\magnification 1200

%
% ??/epsf.tex (written by Radical Eye Software and copied below)
% defines the macro \epsfbox with one argument,
% the encapsulated PostScript file to include.
% Invoking it causes a \vbox with the natural size of the drawing
% to be inserted at the point of invocation.
% Usually figures are meant to be centered and set off, and possibly
% to have a title and/or a figure number. The macros below do that.
% You should assign values that please you to the variables
% \abovefigskip, \belowfigskip, figtitleskip, figtitlefont,...
% AFTER \inputting the present file: \input /u2/kbi/tex/epsfig
% and BEFORE the first invocation of any of the macros below.
% RESET AT YOUR PLEASURE THE VARIABLES AT THE VERY BOTTOM!
%
% For convenience we make a dimension for figures:
\newdimen\FigSize       \FigSize=.9\hsize % alter at your convenience
%
% For a SCALED HORIZONTALLY CENTERED FIGURE use \epsfig.
% First argument is the horizontal width of the figure, given
% in any way TeX can understand.
% The second argument is an encapsulated PostScript file name(filnam.eps).
% Note the mandatory semicolons between arguments in this example!:
% \epsfig .8\hsize; example.ps;
% will put a centered scaled \vbox of width .8\hsize suitably offset
% at the point of invocation
\newskip\abovefigskip   \newskip\belowfigskip
\gdef\epsfig#1;#2;{\par\vskip\abovefigskip\penalty -500
   {\everypar={}\epsfxsize=#1\nd
    \centerline{\epsfbox{#2}}}%
    \vskip\belowfigskip}%
%
% SCALED TITLED EPSFIG HORIZONTALLY CENTERED: \tepsfig.
% First argument is the horizontal width of the figure,
% second an encapsulated PostScript file name,
% third a title for the figure.
% Note the mandatory semicolons between arguments!
% example: \tepsfig5truein; example.ps;{This is a figure}
\newskip\figtitleskip
\gdef\tepsfig#1;#2;#3{\par\vskip\abovefigskip\penalty -500
   {\everypar={}\epsfxsize=#1\nd
    \vbox
      {\centerline{\epsfbox{#2}}\vskip\figtitleskip
       \centerline{\figtitlefont#3}}}%
    \vskip\belowfigskip}%
%
% SCALED NUMBERED TITLED EPSFIG HORIZONTALLY CENTERED: \nepsfig
% The figure number is automatically increased for every
% invocation of \nepsfig or \nipsfig.
\newcount\FigNr \global\FigNr=0
\gdef\nepsfig#1;#2;#3{\global\advance\FigNr by 1
   \tepsfig#1;#2;{Figure\space\the\FigNr.\space#3}}%
%
%
% Often you would rather have TeX decide where to put the figure
% by using \midinsert. Here are macros that do that
% (mnemonics ``ipsfig'' is for ``midInsert PS FIGure'')
%
% TeX-PLACED SCALED EPSFIG HORIZONTALLY CENTERED: \ipsfig
\gdef\ipsfig#1;#2;{%\goodbreak ??
   \midinsert{\everypar={}\epsfxsize=#1\nd
              \centerline{\epsfbox{#2}}}%
   \endinsert}%
%
% TeX-PLACED SCALED TITLED EPSFIG HORIZONTALLY CENTERED: \tipsfig
\gdef\tipsfig#1;#2;#3{\midinsert
   {\everypar={}\epsfxsize=#1\nd
    \vbox{\centerline{\epsfbox{#2}}%
          \vskip\figtitleskip
          \centerline{\figtitlefont#3}}}\endinsert}%
%
% TeX-PLACED SCALED NUMBERED TITLED EPSFIG HORIZONTALLY CENTERED: \nipsfig
% example: \nipsfigd.9\hsize;example.ps;{This is an example figure}
\gdef\nipsfig#1;#2;#3{\global\advance\FigNr by1%
  \tipsfig#1;#2;{Figure\space\the\FigNr.\space#3}}%
\newread\epsffilein    % file to \read
\newif\ifepsffileok    % continue looking for the bounding box?
\newif\ifepsfbbfound   % success?
\newif\ifepsfverbose   % report what you're making?
\newdimen\epsfxsize    % horizontal size after scaling
\newdimen\epsfysize    % vertical size after scaling
\newdimen\epsftsize    % horizontal size before scaling
\newdimen\epsfrsize    % vertical size before scaling
\newdimen\epsftmp      % register for arithmetic manipulation
\newdimen\pspoints     % conversion factor
\pspoints=1bp          % Adobe points are `big'
\epsfxsize=0pt         % Default value, means `use natural size'
\epsfysize=0pt         % ditto
\def\epsfbox#1{\global\def\epsfllx{72}\global\def\epsflly{72}%
   \global\def\epsfurx{540}\global\def\epsfury{720}%
   \def\lbracket{[}\def\testit{#1}\ifx\testit\lbracket
   \let\next=\epsfgetlitbb\else\let\next=\epsfnormal\fi\next{#1}}%
\def\epsfgetlitbb#1#2 #3 #4 #5]#6{\epsfgrab #2 #3 #4 #5 .\\%
   \epsfsetgraph{#6}}%
\def\epsfnormal#1{\epsfgetbb{#1}\epsfsetgraph{#1}}%
\def\epsfgetbb#1{%
%
%   The first thing we need to do is to open the
%   PostScript file, if possible.
%
\openin\epsffilein=#1
\ifeof\epsffilein\errmessage{I couldn't open #1, will ignore it}\else
%
%   Okay, we got it. Now we'll scan lines until we find one that doesn't
%   start with %. We're looking for the bounding box comment.
%
   {\epsffileoktrue \chardef\other=12
    \def\do##1{\catcode`##1=\other}\dospecials \catcode`\ =10
    \loop
       \read\epsffilein to \epsffileline
       \ifeof\epsffilein\epsffileokfalse\else
%
%   We check to see if the first character is a % sign;
%   if not, we stop reading (unless the line was entirely blank);
%   if so, we look further and stop only if the line begins with
%   `%%BoundingBox:'.
%
          \expandafter\epsfaux\epsffileline:. \\%
       \fi
   \ifepsffileok\repeat
   \ifepsfbbfound\else
    \ifepsfverbose\message{No bounding box comment in #1; using
defaults}\fi\fi
   }\closein\epsffilein\fi}%
%
%   Now we have to calculate the scale and offset values to use.
%   First we compute the natural sizes.
%
\def\epsfsetgraph#1{%
   \epsfrsize=\epsfury\pspoints
   \advance\epsfrsize by-\epsflly\pspoints
   \epsftsize=\epsfurx\pspoints
   \advance\epsftsize by-\epsfllx\pspoints
%
%   If `epsfxsize' is 0, we default to the natural size of the picture.
%   Otherwise we scale the graph to be \epsfxsize wide.
%
   \epsfxsize\epsfsize\epsftsize\epsfrsize
   \ifnum\epsfxsize=0 \ifnum\epsfysize=0
      \epsfxsize=\epsftsize \epsfysize=\epsfrsize
%
%   We have a sticky problem here:  TeX doesn't do floating point
arithmetic!
%   Our goal is to compute y = rx/t. The following loop does this reasonably
%   fast, with an error of at most about 16 sp (about 1/4000 pt).
%
     \else\epsftmp=\epsftsize \divide\epsftmp\epsfrsize
       \epsfxsize=\epsfysize \multiply\epsfxsize\epsftmp
       \multiply\epsftmp\epsfrsize \advance\epsftsize-\epsftmp
       \epsftmp=\epsfysize
       \loop \advance\epsftsize\epsftsize \divide\epsftmp 2
       \ifnum\epsftmp>0
          \ifnum\epsftsize<\epsfrsize\else
             \advance\epsftsize-\epsfrsize \advance\epsfxsize\epsftmp
\fi
       \repeat
     \fi
   \else\epsftmp=\epsfrsize \divide\epsftmp\epsftsize
     \epsfysize=\epsfxsize \multiply\epsfysize\epsftmp
     \multiply\epsftmp\epsftsize \advance\epsfrsize-\epsftmp
     \epsftmp=\epsfxsize
     \loop \advance\epsfrsize\epsfrsize \divide\epsftmp 2
     \ifnum\epsftmp>0
        \ifnum\epsfrsize<\epsftsize\else
           \advance\epsfrsize-\epsftsize \advance\epsfysize\epsftmp \fi
     \repeat
   \fi
%
%  Finally, we make the vbox and stick in a \special that dvips can parse.
%
   \ifepsfverbose\message{#1: width=\the\epsfxsize,
height=\the\epsfysize}\fi
   \epsftmp=10\epsfxsize \divide\epsftmp\pspoints
   \vbox to\epsfysize{\vfil\hbox to\epsfxsize{%
      \includegraphics{#1}%
      \hfil}}%
\epsfxsize=0pt\epsfysize=0pt}%
%
%   We still need to define the tricky \epsfaux macro. This requires
%   a couple of magic constants for comparison purposes.
%
{\catcode`\%=12
\global\let\epsfpercent=%\global\def\epsfbblit{%BoundingBox}}%
%
%   So we're ready to check for `%BoundingBox:' and to grab the
%   values if they are found.
%
\long\def\epsfaux#1#2:#3\\{\ifx#1\epsfpercent
   \def\testit{#2}\ifx\testit\epsfbblit
      \epsfgrab #3 . . . \\%
      \epsffileokfalse
      \global\epsfbbfoundtrue
   \fi\else\ifx#1\par\else\epsffileokfalse\fi\fi}%
%
%   Here we grab the values and stuff them in the appropriate definitions.
%
\def\epsfgrab #1 #2 #3 #4 #5\\{%
   \global\def\epsfllx{#1}\ifx\epsfllx\empty
      \epsfgrab #2 #3 #4 #5 .\\\else
   \global\def\epsflly{#2}%
   \global\def\epsfurx{#3}\global\def\epsfury{#4}\fi}%
%
%   We default the epsfsize macro.
%
\def\epsfsize#1#2{\epsfxsize}%
%
%   Finally, another definition for compatibility with older macros.
%

% ================================================================
% execution: why not set
\epsfverbosetrue                        % reset at your pleasure
\abovefigskip=\baselineskip             % reset at your pleasure
\belowfigskip=\baselineskip             % reset at your pleasure
\global\let\figtitlefont\bf             % reset at your pleasure
\global\figtitleskip=.5\baselineskip    % reset at your pleasure

\font\tenmsb=msbm10   %%%% these first lines are to define \Bbb, for R etc.
\font\sevenmsb=msbm7
\font\fivemsb=msbm5
\newfam\msbfam
\textfont\msbfam=\tenmsb
\scriptfont\msbfam=\sevenmsb
\scriptscriptfont\msbfam=\fivemsb

\let\nd\noindent %                                              NOINDENT
%                                          NEWLINE

\def\natural{{\rm I\kern-.18em N}}
 % or scaled\magstep2       TO INSERT LARGE
                            % FONT
%***************  TO GET SMALLER FONT FAMILIES  *****************
\newskip\ttglue

% ********** EIGHT POINT **************

\def\eightpoint{\def\rm{\fam0\eightrm}  
  \textfont0=\eightrm \scriptfont0=\sixrm \scriptscriptfont0=\fiverm
  \textfont1=\eighti  \scriptfont1=\sixi  \scriptscriptfont1=\fivei
  \textfont2=\eightsy  \scriptfont2=\sixsy  \scriptscriptfont2=\fivesy
  \textfont3=\tenex  \scriptfont3=\tenex  \scriptscriptfont3=\tenex
  \textfont\itfam=\eightit  \def\it{\fam\itfam\eightit}
  \textfont\slfam=\eightsl  \def\sl{\fam\slfam\eightsl}
  \textfont\ttfam=\eighttt  \def\tt{\fam\ttfam\eighttt}
  \textfont\bffam=\eightbf  \scriptfont\bffam=\sixbf
    \scriptscriptfont\bffam=\fivebf  \def\bf{\fam\bffam\eightbf}
  \tt  \ttglue=.5em plus.25em minus.15em
  \normalbaselineskip=9pt
  \setbox\strutbox=\hbox{\vrule height7pt depth2pt width0pt}
  \let\sc=\sixrm  \let\big=\eightbig \normalbaselines\rm}

\font\eightrm=cmr8 \font\sixrm=cmr6 \font\fiverm=cmr5
\font\eighti=cmmi8  \font\sixi=cmmi6   \font\fivei=cmmi5
\font\eightsy=cmsy8  \font\sixsy=cmsy6 \font\fivesy=cmsy5
\font\eightit=cmti8  \font\eightsl=cmsl8  \font\eighttt=cmtt8
\font\eightbf=cmbx8  \font\sixbf=cmbx6 \font\fivebf=cmbx5

\def\eightbig#1{{\hbox{$\textfont0=\ninerm\textfont2=\ninesy
        \left#1\vbox to6.5pt{}\right.\enspace$}}}

%************** NINE POINT *****************
\def\ninepoint{\def\rm{\fam0\ninerm}  
  \textfont0=\ninerm \scriptfont0=\sixrm \scriptscriptfont0=\fiverm
  \textfont1=\ninei  \scriptfont1=\sixi  \scriptscriptfont1=\fivei
  \textfont2=\ninesy  \scriptfont2=\sixsy  \scriptscriptfont2=\fivesy
  \textfont3=\tenex  \scriptfont3=\tenex  \scriptscriptfont3=\tenex
  \textfont\itfam=\nineit  \def\it{\fam\itfam\nineit}
  \textfont\slfam=\ninesl  \def\sl{\fam\slfam\ninesl}
  \textfont\ttfam=\ninett  \def\tt{\fam\ttfam\ninett}
  \textfont\bffam=\ninebf  \scriptfont\bffam=\sixbf
    \scriptscriptfont\bffam=\fivebf  \def\bf{\fam\bffam\ninebf}
  \tt  \ttglue=.5em plus.25em minus.15em
  \normalbaselineskip=11pt
  \setbox\strutbox=\hbox{\vrule height8pt depth3pt width0pt}
  \let\sc=\sevenrm  \let\big=\ninebig \normalbaselines\rm}

\font\ninerm=cmr9 \font\sixrm=cmr6 \font\fiverm=cmr5
\font\ninei=cmmi9  \font\sixi=cmmi6   \font\fivei=cmmi5
\font\ninesy=cmsy9  \font\sixsy=cmsy6 \font\fivesy=cmsy5
\font\nineit=cmti9  \font\ninesl=cmsl9  \font\ninett=cmtt9
\font\ninebf=cmbx9  \font\sixbf=cmbx6 \font\fivebf=cmbx5
\def\ninebig#1{{\hbox{$\textfont0=\tenrm\textfont2=\tensy
        \left#1\vbox to7.25pt{}\right.$}}}
%%%%%%%%%%%%%%%%%%%%%%%%%%%%%%%%%%%%%%%%%%%%%%%%%%%%%%%%%%%%%%%%%%%%%%%%%%%

\def\chix{{\raise.5ex\hbox{$\chi$}}}
\def\chixa{{\chix\lower.2em\hbox{$_A$}}}

\def\real{{\rm I\kern-.2em R}}
\def\integer{{\rm Z\kern-.32em Z}}
\def\complex{\kern.1em{\raise.47ex\hbox{
            $\scriptscriptstyle |$}}\kern-.40em{\rm C}}
\def\vs#1 {\vskip#1truein}
\def\hs#1 {\hskip#1truein}
  \hsize=6truein \hoffset=.2truein %was \hoffset=1.2truein
  \vsize=8.8truein %\voffset=1truein 
\pageno=1 \baselineskip=12pt
  \parskip=0 pt \parindent=20pt 
\overfullrule=0pt \lineskip=0pt \lineskiplimit=0pt
  \hbadness=10000 \vbadness=10000 % REPORT ONLY BEYOND THIS BADNESS 
     \pageno=0
     
     \footline{\ifnum\pageno=0\hss\else\hss\tenrm\folio\hss\fi}
     \hbox{}
     \vskip 1truein\centerline{{\bf Random Loose Packing in Granular Matter}}
     \vskip .2truein\centerline{by}
     \vskip .2truein
\centerline{{David Aristoff}
\ \ and\ \  {Charles Radin}
\footnote{*}{Research supported in part by NSF Grant DMS-0700120\hfil}}

     \vskip .1truein\centerline{
     Mathematics Department, University of Texas, Austin, TX 78712}
     \vs.2
\vs.5
     \centerline{{\bf Abstract}}
     \vs.1 \nd
     We introduce and simulate a two dimensional probabilistic model
     of granular matter at vanishing pressure. The model exhibits a
     perfectly sharp random loose packing density, a phenomenon that
     should be verifiable for real granular matter.

\vs3
\centerline{August, 2008}
\vs1
\centerline{PACS Classification:\ \ 45.70.Cc, 81.05.Rm, 05.70.Ce}
     \vfill\eject

\nd {\bf 1. Introduction} \vs.1 

We introduce and analyze a crude model for the random loose packings
of granular matter. These packings, as well as random close packings,
were carefully prepared by Scott et al in the 1960's [S,SK], mainly
in samples of steel ball bearings. Gently pouring samples of 
20,000 to 80,000 spheres into a container, the lowest possible volume
fraction obtainable -- the so-called random loose packing density -- was
determined to be $0.608\pm 0.006$.

The above refers to monodisperse steel spheres immersed in air;
similar experiments were performed with spheres of other materials 
immersed in other fluids; variations in the coefficient of
friction and in the effective gravitational force lead to somewhat
different values for the random loose packing density [SK].

Matter is generally described as ``granular'' if it is composed of a
large number of noncohesive subunits each of which is sufficiently
massive that its gravitational energy is much larger than its thermal
energy. A common example is sand.

There are several classic phenomena characteristic of static granular
matter, in particular dilatancy, random close packing, and random
loose packing, none of which can yet be considered well-understood;
see [dG] for a good review. A basic question about these
phenomena is whether they are sharply defined or inherently
vague. Dilatancy has recently been associated with a phase transition
measured by the response of the material to shear [SN], which answers
the question for this phenomenon. The case of random close packing is
controversial and awaits further experiment; see [TT,R]. Our main goal
here is to analyze this question with respect to random loose packing, to
determine whether or not traditional theoretical approaches to
granular matter predict a sharply defined random loose packing
density. It is clear that any experimental determination of a random
loose packing density will vary with physical conditions such as
coefficient of friction. It is possible, however, that there is a
precise geometrical quantity which underlies the phenomenon, and it is
our goal to investigate this question.

We will eventually specialize to a certain two dimensional model, but
we begin our discussion more ambitiously.  Consider a model using a
large collection of impenetrable, unit mass, unit diameter spheres in a large
container, acted on by gravity and with infinite coefficient of
friction between themselves and with the container.  We will not try
to understand a random loose packing density as an {\it absolute
minimum} at which the spheres can form a bulk material. Rather, we
will use a probabilistic framework in the spirit of Edwards' theory of
granular matter [EO].  More specifically, we expect, but cannot show,
that the following is true, and use it as our motivation. Consider a
probability density on the set of all mechanically stable packings of
the spheres in their container, with the probability density of a
packing $c$ proportional to $\exp[-E(c)]$, where $E(c)$ is the sum of
the heights, from the floor of the container, of the centers of the
spheres in the packing $c$. We expect that such an ensemble will
exhibit a gradient in the volume fraction, with volume fraction
decreasing with height, and that there is a well-defined random loose
packing density as one approaches the top of the packing, where the
(analogue of hydrostatic) pressure goes to zero. More specifically, we
expect that as one takes an infinite volume limit, the 
probability distribution for the volume fraction of the top
layer of the packing becomes concentrated at a single nonzero
value. We emphasize that we are focusing on a bulk property near the
top of the configuration, not a surface phenomenon.

To analyze random loose packing in such a probabilistic framework we
make several simplifications of the above model, first to an ensemble
of those packings which are limits, as the gravitational constant goes
to zero, of mechanically stable packings; we effect this by setting
$E(c)=0$ in the relative density $\exp[-E(c)]$. With this simplification
configurations are now, in their entirety, representative of the top 
layer in the original model. Next we consider the two
dimensional version of the above: congruent frictional unit disks in
mechanically stable configurations under vanishingly small
gravity. Note that each such disk must be in contact with either a
pair of supporting disks below it or part of the
container. (Here and elsewhere in this paper we neglect events of
probability zero, such as one sphere perfectly balanced on another.)
We simplify the model one last time by replacing the disks by
congruent squares, with edges aligned with the sides of the
(rectangular) container, each square in contact with either a pair of
supporting squares below it or the floor of the container. This is
now a granular version of the old model of ``(equilibrium) hard
squares'' [H], which is a simplification of ``hard disks'' and ``hard
spheres'' (see [AH] for a review), in which gravity is neglected but kinetic 
energy plays a
significant role. We emphasize that in our granular model there is no
longer any need to concentrate on the ``top layer''; in fact we will
eventually be concerned with an infinite volume limit which, as
usual, focuses on the middle of the collection of squares 
and lets the boundaries grow to infinity.

It is this granular model which is the focus of this
paper.  We have run Markov chain Monte Carlo simulations on the model,
with the following results. We initialize the squares in an allowed
configuration of some well-defined volume fraction anywhere between
0.5 and 1. If the initial volume fraction $\phi$ is not approximately
$0.76$, the packing expands or contracts, relatively quickly, into
the range $0.76\pm 0.01$; see Figs.\ 1 and 2. 

The process is
insensitive to the tightness of fit of the initial configuration in
the containing box except for absolute extremes. If the side walls of
the containing box abut a closely-packed initial configuration, the simulation
cannot
significantly change the volume fraction; alternatively
if the width of the box is much larger than that of the initial configuration
the simulation will drive the configuration to a monolayer on the
floor; but both extremes are negligible and the equilibrium volume
fraction is otherwise insensitive to the fit of the initial configuration in
the box. More precisely, we found that the equilibrium volume fraction should 
be accurate if the floor length is between $2 \sqrt{N}$ and $8 \sqrt{N}$, where 
$N\ge 100$ is the number of squares. Since we will be conjecturing the
behavior of the model in the infinite volume limit, the
equilibrium configuration should be a single bulk pile, so the floor 
length should be on the order of $\sqrt{N}$. To understand 
the lower bound, note that, at any volume fraction, a configuration occupies the 
least amount of floor space when the squares are arranged in a single full triangle. 
The bottom level of such a triangle has just under $\sqrt{2N}$ squares. 
Assume the containing box fits tightly around the triangle. If the triangle
has volume fraction greater than $0.754$ then the configuration will not be able to 
decrease to this equilibrium volume fraction; we avoid this by ensuring that the floor 
length is at least $(0.754)^{-1}\sqrt{2N}$. To arrive at the upper bound we 
performed simulations on fixed particle number and let the floor length vary. We 
saw that the equilibrium volume fraction was reliable so long as the floor length was less 
than about $8\sqrt{N}$, at least for $N\ge 100$.

We conclude that, for floor lengths in the aforementioned acceptable range, 
the equilibrium volume fraction
is found to depend only on the number of squares in the system. The main goal of our
work is an analysis of the distribution of volume fraction -- both the
mean and standard deviation -- as the
number of particles increases. We conclude that the limiting standard
deviation as particle number goes to infinity is zero, so the model
exhibits a sharp value for the random loose packing density, which we
estimate to be approximately $0.754$.

The heart of our argument is the degree to which we can demonstrate
that in this model there is a sharp value, approximately $0.754$, for
the equilibrium volume fraction of large systems, and we postpone 
analysis of error bars to later sections. But to understand the
value $0.754$, consider the following crude estimate of the volume in
phase space of all allowable packings at fixed volume fraction
$\phi$. First notice that the conditions of the model prevent the
possibility of any ``holes'' in a configuration. Furthermore, if we
consider any rectangle in the interior of a configuration, each
horizontal row in the rectangle contains the same number of
squares. (One consequence is that in the infinite volume limit 
each individual configuration must have a sharply defined volume fraction; of
course this is quite distinct from the degree of spread of the volume fraction
among all configurations.) Now consider a very
symmetrical configuration of squares at any desired volume fraction
$\phi$, with the squares in each horizontal row equally spaced, and
gaps between squares each of size $(1-\phi)/\phi$ centered over
squares in the next lower horizontal row; see Fig.$\ 3$. Consider these
squares to represent average positions, fix all but one square in such a
position, and consider the (horizontal) degree of motion allowed to
the remaining square. There are two constraints on its movement: the
gap size separating it from its two neighbors in its horizontal row,
and the length to which its top edge and bottom edge intersects the
squares in the horizontal rows above and below it. These two
constraints are to opposite effect: increasing the gap size decreases
the necessary support in the rows above and below. A simple
calculation shows that the square has optimum allowed motion when the
gap size is $1/3$, corresponding to a
volume fraction of $0.75$, roughly as found in the simulations. In other words,
this crude estimate suggests that the volume in phase space (which we
effectively estimate, for $N$ squares, to be $L^N$ where $L$ is the
allowed degree of motion of one square considered above) is
maximized among allowed packings of fixed volume fraction by the
packings of volume fraction about $0.75$.

To obtain accurate physical measurements a method of sedimentation has
been developed to prepare samples of millions of grains in a
controlled manner; see [JS] and references therein for the current
state of the experimental data. In these experiments monodisperse
grains sediment in a fluid. The sediment is of uniform volume
fraction, at or above 0.55 depending on various experimental
parameters. To achieve the low value ($0.55 \pm 0.001$ [JS]), the
grains need to have a high friction coefficient and the fluid needs
to have mass density only slightly lower than the grains, to minimize
the destabilizing effect of gravity. (In the absence of gravity one
could still produce a granular bed by pressure; we do not know of
experiments reporting a random loose packing value for such an
environment.)

Our results suggest that whatever the initial local volume fraction of the
bed, on sedimentation (in low effective gravity) most samples would
have a well-defined volume fraction, the random loose packing density of
about 0.55, with {\it no intrinsic lower bound} on the sharpness of
the value. This should be verifiable by a sequence of experiments of
increasing dimensions.

There have been previous probabilistic interpretations of the random loose
packing density, for instance [MP,PC]. A distinguishing feature of our results is
our analysis of the degree of sharpness of the basic notion, which, as
we shall see below, requires unusual care in the treatment of error
analysis.

\vs.2
\nd {\bf 2. Results} \vs.1 

We performed Markov chain Monte Carlo simulations on our granular
model, which we now describe more precisely.  We begin with a
fixed number of unit edge squares contained in a large rectangular box
$B$. A collection of squares is ``allowed'' if they do not overlap
with positive area, their edges are parallel to those of the box $B$,
and the lower edge of each square intersects either the floor of the
box $B$ or the upper edge of each of two other squares;
see Fig.\ 3. Note that although the squares have continuous
translational degrees of freedom in the horizontal direction, 
this is not in evidence in the
vertical direction because of the stability condition: the squares
inevitably appear at discrete horizontal ``levels''.

  Markov chain simulations were performed as follows. In the
  rectangular container $B$ a fixed number of squares are introduced
  in a simple ``crystalline'' configuration: squares are arranged
  equally spaced in horizontal rows, the spacing determined by a
  preassigned volume fraction $\phi$, and with squares centered above
  the centers of the gaps in the row below it; see Fig.\ 4. The basic
  step in the simulation is the following. A square is chosen at
  random from the current configuration and all possible positions are
  determined to which it may be relocated and produce an allowed
  configuration. Note that if the chosen square supports a square
  above it then it can only be allowed a relatively small horizontal
  motion; otherwise it may be placed atop some pair of squares, or the
  floor. So the boundary of the configuration plays a crucial role in
  the ability of the chain to change the volume fraction. 
  In any case the positions to which the chosen square may be
  moved constitute a union of intervals. A random point is selected
  from this union of intervals and the square is moved. The random
  movement of a random square is the basic element of the Markov
  chain. It is easy to see that this protocol is transitive and
  satisfies detailed balance, so the chain has the desired uniform
  probability distribution as its asymptotic state [NB]. See Fig.\ 5
  for a configuration of 399 squares after $10^6$ moves. Our
  interest is in random loose packing, which occurs in the top (bulk) layer
  of a granular pile, and we assume that the entirety of each of our configurations 
  represents this top layer. We emphasize that our protocol is
  not particularly appropriate for studying other questions such as
  the statistical shape of the boundary of a granular pile, or properties
  associated with high volume fraction, such as random close packing.  

After a prescribed number of moves, a volume fraction is computed for
the collection of squares as follows. Within horizontal level $L_j$,
where $j=0$ corresponds to the squares resting on the floor, the
distances between the centers of neighboring squares is
computed. (Such a distance is $1+g$ where $g$ is the gap between the squares.)
Suppose that $n_j$ of these neighboring distances are each less than 2, and
that the sum of these distances in the level is $s_j$. At this point
our procedure will be
complicated by the desire to obtain information during the simulation
about inhomogeneities in the collection, for later use in analyzing
the approach to equilibrium. For this purpose we introduce a new
parameter, $p$.
For fixed $0< p < 1$ we consider
those levels, beginning from $j=0$, for which $n_j$ is at least
$0.75p$ times the length of the box's floor. Suppose $L_{J(p)}$ is the highest 
level such that it, and all levels below it, satisfy the condition. We then assign the
volume fraction

$$\phi(p)={\sum_{j=0}^{J(p)} n_j\over \sum_{j=0}^{J(p)} s_j} \eqno{(1)}$$

\nd to the assembly of squares. (The factor $0.75$ 
represents the volume fraction we expect the box's floor to reach in equilibrium.
 Note that
any two such calculations
of volume fraction of the same configuration may only differ by a
term proportional to the length of the boundary of the configuration,
so any inhomogeneity is limited to this size.)
Such a calculation of volume fraction was performed regularly, after approximately
$10^{6}$ moves, producing a time series of volume fractions $\phi_t$
for the given number of squares. (We suppress reference to the
variable $p$ for ease of reading. As will be seen later all our
results correspond to the choice $p=0.4$, so one can, without much loss, ignore other
possibile values.) Variables $\phi_t$ and $\phi_{t+1}$
are highly dependent, but we can be guaranteed that if the series
is long enough then the sample mean:

$${1\over N} \sum_{t=1}^N  \phi_t \eqno{(2)}$$

\nd will be a good approximation to the true mean of the target
(uniform) probability distribution for the given number of squares [KV].

We created such time series $\phi_t$, each of about $10^4$ terms
(roughly $10^{10}$ elementary moves),
using values $p=0.2,0.4,0.6,0.8$, on systems for the following numbers
of squares: $100m$, and $1000m$, for $m=1,2,\ldots,9$, with varying
initial volume fractions. For each system
we needed to determine the initialization period -- the number of
moves necessary to reach equilibrium -- and then the total number of
moves to be performed. Both of these determinations were made based on
variants of the (sample) autocorrelation function $f(k)$ of the time
series $\{\phi_t\ |\ 1\le t\le T\}$ of volume fractions, defined for
$1\le k\le T$ by:

$$f(k)= {1\over T-k+1}\sum_{t=1}^{T-k+1}
(\phi_t-\bar{\phi})(\phi_{t+k}-\bar{\phi}), \eqno{(3)}$$

\nd where $\bar{\phi}$ is the mean of the series.
This function is easily seen to
give less reliable results as $k$ increases, because of limited data, so 
it is usual to work with
functions made from it as follows. One way to avoid difficulties
is to restrict the domain of $f$; we define 
the ``unbiased'' autocorrelation
$f_1(k)$ by
$f_1(k)=f(k)$ for $k\le T/10$. Another variant we consider
is the ``biased'' autocorrelation $f_2$, defined for all $1\le
k\le T$ by:

$$f_2(k)= {1\over T}\sum_{t=1}^{T-k+1}
(\phi_t-\bar{\phi})(\phi_{t+k}-\bar{\phi}), \eqno{(4)}$$

\nd which reduces the value of $f(k)$ for large $k$. (See pages
321-324 of Priestley [P] for a discussion of this biased variant.)
We consider both variants of autocorrelation; to refer to either we
use the term $f_j$.

With these autocorrelations we determined the smallest $k=k_z$ such that
$f_j(k)=0$. We then computed the sample standard
deviation $\sigma_{f_j}$ away from zero of $f_j$ restricted to $k\ge k_z$, and defined
$k_I$ to be the smallest $k$ such that $|f_j(k)|\le \sigma_{f_j}$; see Fig.\ 6.
(For ease of reading we sometimes do not add reference to $j$ to quantities
derived using $f_j$.)
This defined the initialization period. Then starting from
$\phi_{k_I}$ we recomputed the autocorrelation $\tilde f_j$ and 
$\sigma_{\tilde f_j}$ and determined the mixing
time, the smallest $k=k_M$ such that $|\tilde f_j(k)|
\le \sigma_{\tilde f_j}$. $k_M$ was interpreted as the separation $k$
needed such that the random variables $\phi_t$ and $\phi_{t+k}$ are
roughly independent for all $t\ge k$. (We performed the above using
the different definitions of volume fraction corresponding to
different values of $p$, allowing us to analyze different geometrical
regions of the samples. For each system of squares we selected, for
initialization and mixing times, the largest obtained as above
corresponding to the various values of $p$, which was always that for
$p=0.2$, corresponding to the lowest layers of the configuration.)

Once we determined $k_I$ and $k_M$ we ran the series to $\phi_F$, where
$F=k_I+Tk_M$ for some $T\ge 20$.  The values of $k_I$ and $k_M$ are
given in Tables 1 and 2; the empirical means and
standard deviations of volume fraction are given in Tables 3 and 4.

In all our results we use $p=0.4$ to minimize
the boundary effects presumably associated with small or large $p$. 
(With large $p$ the lowest level may have undue influence on the 
volume fraction; with small $p$, the surface levels could have undue influence. 
Note that the arrangements of squares on the lowest level and the
surface levels are not restricted by the  arrangements of squares
below and above them, respectively, and so the corresponding volume fractions 
are not bound to the logic, discussed above, which suggested that each level 
should equilibrate 
at a volume fraction of about .75. In spite of this, we found that using 
any series corresponding to $p$ in the range $0.2 \le p \le 0.8$ 
generated a similar result.)  For all 
systems the volume fraction quickly settles to the range $0.76\pm
0.01$
and we can easily see from Table 4
that the empirical standard deviations decrease with increase of
particle number. In Fig.\ 7 we plot the empirical standard
deviations against particle number, and in Fig.\ 8 the
data is replotted using logarithmic scales. In Fig.\ 8 we include 
the best least-squares fit to a straight line $y=ax+b$, obtaining
$a=-0.5004$ and $b=-0.8052$. The corresponding curve is included in
Fig.\ 7. Also included in both graphs are $90\%$ confidence
intervals for the true standard deviations, obtained as
described in the next section, using $f_2$. (There was not enough data 
to obtain a confidence interval by this method for the system with 8995 squares.)
The same data is
reanalyzed in Figs.\ 9 and 10 with confidence intervals derived using $f_1$.

We use the close fit to the line in Figs.\ 8 or 10, corresponding to
33 data points in a range of particle number varying from 100 to 9000, to extend the
agreement to arbitrarily large particle number, and therefore to claim
that the standard deviation is zero in the infinite volume limit, or
that there is a sharp value for the random loose packing density. The
argument is supported from the theoretical side by noting the closeness of the slopes
in Figs.\ 8 and 10 to $-1/2$. A slope of $-1/2$ would be expected if it were true that
an equilibrium configuration of $N$ squares could be partitioned into similar subblocks
which are roughly independent -- a proposition which would not be surprising given a
phase interpretation of granular media [R]. Verifying such independence
might be of some independent interest but would require much more
data and much longer running times.

This is our main result, since it shows how to make sense of a
perfectly well-defined random loose packing density within a granular
model of the standard Edwards' form.

As to the actual asymptotic value of the volume fraction in the limit of large
systems, we assume that our simulations suffer from a surface error
proportional to $\sqrt{N}$ for a system of $N$ squares. The least-squares
fit of a function of form $\displaystyle A + B/\sqrt{N}$ to the
data (see Figs.\ 11 and 12) 
yields $A=0.7541$, and the good fit suggests an (asymptotic)
random loose packing density in our granular model of about $0.754$.

Our argument concerning asymptotically large systems depends on the
fit of our standard deviation data to a curve, and the degree to which this fit is
convincing depends on the confidence intervals associated with our simulations.
In the next section we explain how we arrived at our confidence
intervals.

\vs.2
\nd {\bf 3. Simulation data analysis} \vs.1 

A good source for common ways to analyze the data in Markov chain
Monte Carlo simulation is chapter 3 in Newman and Barkema [NB]. We will
give a more detailed analysis, following the paper by Geyer [Gey] in the
series put together for this purpose by the statistics community [Gel].
As will be seen, our argument is based on the precision of estimates
of various statistical quantities, and necessitates a delicate treatment.

Our simulations produce a time series $c_j$ of (dependent) random
configurations of squares. From this we produce other series $g(c_j)$
using functions $g$ on the space of possible configurations $c$, in
particular the volume fraction $g_1(c)=\phi$ and $g_2(c)=(\phi-K)^2$
for constant $K$.

We use the common method of batch means. As described in the previous
section, we first
determine an initialization time $k_I$ and a mixing time $k_M$ for our
series $c_j$ from autocorrelations. After removing the initialization
portion of the series, we break up the remaining $W$
terms of the series into $w\ge 2$ equal size consecutive batches (subintervals),
each of the same length $W/w$, discarding the last few terms from the
series if $w$ does not divide $W$ evenly.

It should be emphasized that rarely, if ever, 
are conclusions drawn from a finite number of
Monte Carlo simulations a literal proof of anything interesting. We
are going to obtain confidence intervals (using the Student's t-test) 
for the mean and standard
deviation of the volume fraction of our systems of fixed particle
number. The t-test's results would be mathematically rigorous if 
in our simulations we had performed infinitely many moves; of course 
this is impossible, so we will try to make a convincing case
that we have enough data to give reliable results. Ultimately, this
is the most sensitive point in our argument.

Assume fixed some function $g$, and denote the true mean of $g(c)$ by
$\mu_g$.  Assume, temporarily, that enough moves have been taken for
the t-test to be reliable. (We will come back to this assumption below.)
With the notation $\overline {g(c)}$ for
the empirical average $(1/w) \sum_k \langle g(c)\rangle_k$ of $g(c)$,
the variable:

$${ \overline {g(c)} -  \mu_g\over 
  \sqrt{ {1\over w(w-1)}\sum_k (\langle g(c)\rangle_k - \overline
  {g(c)})^2}} \eqno{(5)}$$

\nd has a
t-distribution, allowing one to compute
confidence intervals for $\mu_g$.

The above outline explains how (given the validity of the t-test) 
we could compute confidence intervals for the mean value
of the volume fraction for the time series associated with our
simulations for fixed numbers of squares. A small variation
allows us to give confidence intervals for the standard deviations of
these variables, as follows.

Denote the true 
standard deviation of $g(c)$ by $\sigma_g$. Using conditioning, 

$$\eqalign{\hbox{Prob}(& \mu_g \in I \hbox{ and } \sigma_g \in J)=\hbox{Prob}(\mu_g \in
I)\hbox{Prob}(\sigma_g \in J\ |\ \mu_g \in I)\cr
&=\hbox{Prob}(\mu_g \in
I)\sum_i\hbox{Prob}(\sigma_g \in J\ |\ \mu_g \in I_i)\hbox{Prob}(\mu_g \in
I_i \ |\ \mu_g \in I),\cr} \eqno{(6)}$$

\nd where $\{I_i\}$ is a partition of $I$.
We have discussed how to obtain $I$ so that the factor
$\hbox{Prob}(\mu_g \in I)$ is at least $0.95$. We now want to
obtain $J$ so that the factor $\hbox{Prob}(\sigma_g \in J\ |\
\mu_g \in I)$ is also at least $0.95$, and therefore 
$\hbox{Prob}(\mu_g \in I \hbox{ and } \sigma_g \in J)$ is at least $(0.95)(0.95)>0.90$.

Consider, for each constant $K$, the random variable 
$$\Sigma_K=\sqrt{(1/w) \sum_j
[\langle g(c)\rangle_j - K]^2}. \eqno{(7)}$$ 

\nd Using (5) with $(g(c)-K)^2$ playing the role of $g(c)$, 
we can obtain a $95\%$ confidence
interval for the mean of $\Sigma_K^2$, which we translate into 
a $95\%$ confidence interval $J_K$ for the mean of $\Sigma_K$.
Assume the
partition so fine that within the desired precision $J_K=J_i$ only
depends on $i$, where $K\in I_i$. 
Note that if
$K=\mu_g$, then the random variable $\Sigma_K$ has as its mean the
standard deviation
$\sigma_g$. So if we let $J=\cup_{i} J_i$, then 
$\hbox{Prob}(\sigma_g \in J\ |\ \mu_g \in I_i)>0.95$ for all $i$,
and therefore
$\hbox{Prob}(\sigma_g \in J\ |\ \mu_g \in I)>0.95$. In practice the union $J=\cup_{i} J_i$ is
easy to compute.

In the above arguments we have assumed that enough moves have been
taken to justify the t-test, which has independence and normality
assumptions which are not strictly satisfied in our situation. 
We now consider how to deal with this situation. 
Some guidance 
concerning independence can be obtained from the following toy model.

Assume that for the time series of the simulation one can determine
some number $k_M$, perhaps but not necessarily derived as above from the
autocorrelation $f(k)$, such that variables $\phi_i$ and $\phi_{i+k}$ in the time
series are roughly independent if $k\ge k_M$. We model this
transition between independent random variables as follows.

Let $T$ and $M$ be nonnegative (integer) constants.
For $0\le t\le T$ and $1\le m \le M-1$ we first define independent,
identically distributed random
variables $X_{tM}$ and from these define:

$$X_{tM+m}=\Big( 1-{m\over M}\Big)X_{tM} + {m\over M} X_{(t+1)M},\eqno{(8)}$$

\nd together defining $X_t$ for $0\le t \le TM-1$. Note that variables $X_t$
and $X_{t+m}$ are independent for $m\ge 2M-1$.

A simple calculation shows that:

$$\sum_{m=0}^{M-1} X_{tM+m} = \Big({M+1\over 2}\Big)X_{tM} +
\Big({M-1\over 2}\Big)X_{(t+1)M}.\eqno{(9)}$$

\nd Then another simple calculation shows that:

$$ S_T\equiv {1\over TM}\sum_{m=0}^{TM-1} X_m = {1\over
  T}\Big[\sum_{t=1}^{T-1} X_{tM}+{1\over 2}(X_0+X_{TM}) +{1\over
  2M}(X_0-X_{TM})\Big].\eqno{(10)}$$
\vs0 \nd
In other words $S_T$ is the mean of roughly $K$ 
{\it independent} variables.

Returning to the question of the assumptions in the t-test, the toy
model suggests that the independence assumption is easily
satisfied. The normality assumption is usually taken as the more
serious [Gey]. But we note from [DL] that the t-test is quite robust
with respect to the normality assumption. Although the robustness of
the t-test is well known and is generally relied on, in practice one
still has to pick specific batch partitions in a reliable way. This is not
covered in [Gey]. We arrived at
a standard for batches of length 10 times mixing time for our series as
follows. In outline, we use mixing times as computed above to
standardize comparison between our systems with different particle
number. Those for which our runs constituted at least 800 mixing times
are assumed to give accurate values for the mean volume
fraction. Various initial segments of these runs are then used, with
various choices of batch partitions, to see which choices (if any) give
reliable results for confidence intervals. Batches of size 10 mixing
times proved reliable even for initial segments in the range of 20-100
mixing times, so this choice was then used for all systems. We
emphasize that we are using this method to determine a minimum
reliable batch size
on the sequence of configurations, and then we apply this to the time
series $\phi_t$ as well the time series $[\phi_t-K]^2$. We now
give more details.

For most of the systems of particle numbers 100-900 we have over 500
mixing times worth of data, yet for some of the systems of particle
numbers 1000-9000 we have, for practical reasons, less than one tenth
that depth of data. We want to choose a fixed multiple of mixing time
as batch length for all of our batches. To decide what range of mixing
times will be reliable we used various portions of the data from those of
our longest runs, and then applied the conclusions we drew to the
other 3/4 of the runs.

More specifically, we treat as ``reliable'' the empirical volume
fraction of the longest runs, those of length at least 800 times mixing
time. We then consider a range of batch partitions
of these systems to see which ones give accurate t-test results.
We are looking for $95\%$ confidence
intervals, so we expect such intervals to contain the true volume
fraction $95\%$ of the time; since the true volume fraction is unknown
we instead check how frequently the intervals contain the empirical volume 
fraction, which for the longer runs we have assumed is reliable. 
We do this for each of the runs of length 800 or more times mixing time.
The results on these systems are the following.

For each of our longer runs (of at least 800 mixing times), we considered 
various initial 
portions of the run in each of six ranges of
mixing times: 20-100, 100-200, 300-400, 400-500, and 500-600. For each
of these truncated runs we considered batch partitions of the data into equal
size batches of a variety of multiples of mixing time: 1-5, 6-10, 11-15, 16-20, 
21-30, 31-40 and 41-50. For each size
run and for each batch size we computed a $95\%$ confidence interval for the
true mean of the volume fraction, and determined whether or not the 
confidence interval covers the sample mean for the full run 
(which we are assuming is interchangeable
with the true mean). The fraction of the more than 200 cases in each category for
which the sample mean lies within the confidence interval is recorded
in Table 5. From this it appears that using batches of size 1-5
mixing times would be unreliable, but that size 10 times mixing times
would be reliable. (Table 5 is based on mixing times obtained using
the autocorrelation $f_2$. Table 6 is similar, using the
autocorrelation $f_1$, and again justifies the use of batches of size
10 times mixing time.)

We then used batches of size 10 times mixing times to obtain $95\%$
confidence intervals for the true mean of all the systems,
obtaining the results tabulated in Table 3 and included in Figs.\ 17
and 18.

Finally, we applied the above batch criterion to obtain $90\%$ confidence
intervals for the true standard deviation of all our systems, using the
method described earlier in this section. The results are in Table 4
and in Figs.\ 7 to 10.

\vs.2 \nd
{\bf 4. Conclusion} \vs.1

We have performed Markov chain Monte Carlo simulations on a two
dimensional model of low pressure granular matter of the general
Edwards probabilistic type [EO]. Our main result, superficially
summarized in Fig.\ 8, is that in this model the standard deviation
of the volume fraction decays to zero as the particle number
increases, which indicates a well-defined random loose packing density
for the model. This suggests that real granular
matter exhibits sharply defined random loose packing; this could be
verified by repeating the sedimentation experiments [JS] at a range
of physical dimensions.
Our argument is only convincing to the extent that the
confidence intervals in Fig.\ 8 are small and justified, which
required a statistical treatment of the data unusual in the physics
literature. We hope that our detailed error analysis may be useful in
other contexts. 

\vfill \eject
%zzz
%\end

%\end

\hbox{}
\vs-.3
\centerline{\bf Figures and Tables}
\vs-.1 \nd
\epsfig .9\hsize;  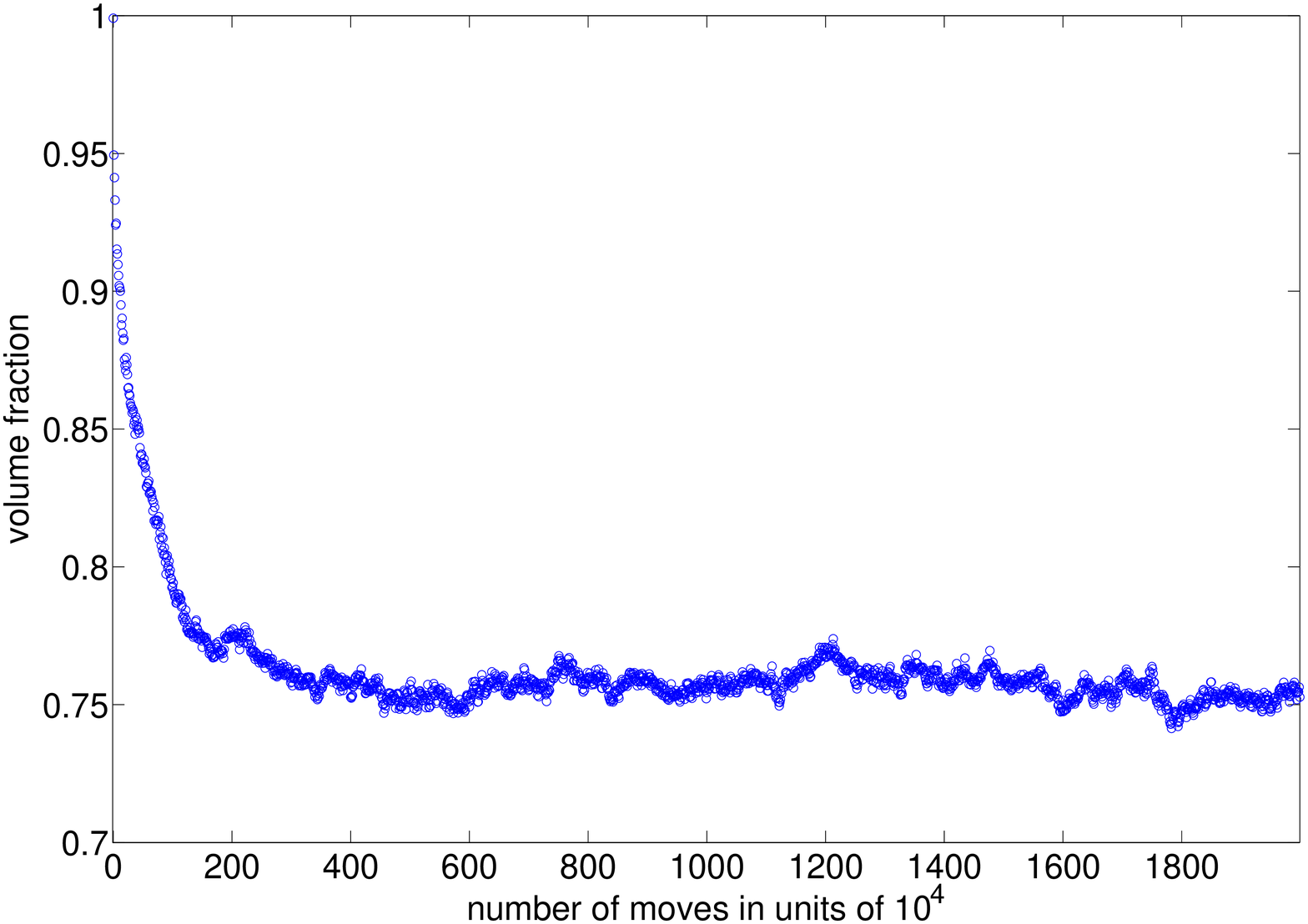;
\vs-.1 \nd
Figure 1. Plot of volume fraction versus number of moves, 
from an initial volume fraction of $0.9991$, for
970 squares.

\vs0 
\hbox{}
\vs0 \nd
\epsfig .9\hsize;  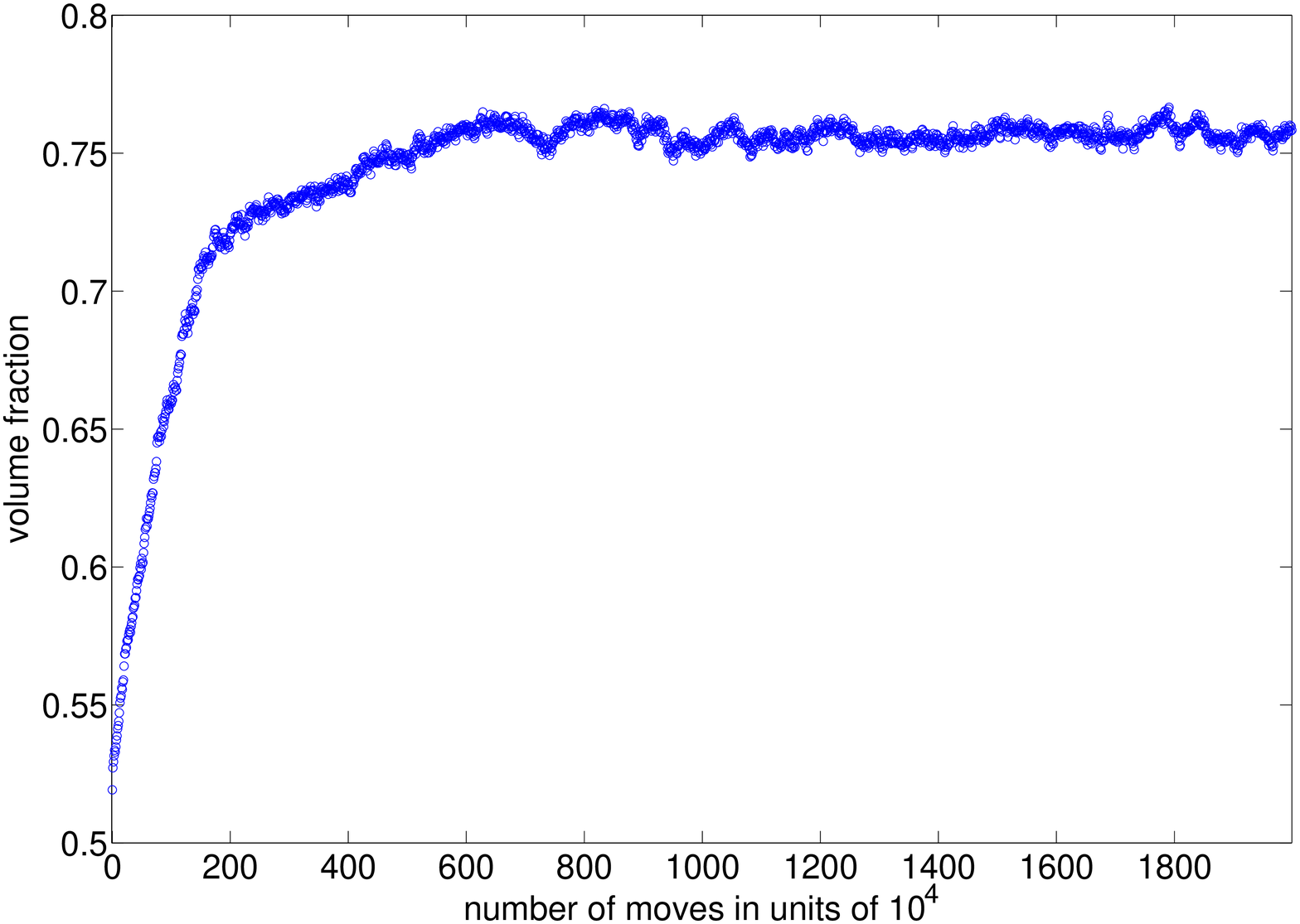;
\vs-.1 \nd
Figure 2. Plot of volume fraction versus number of moves, 
from an initial volume fraction of $0.5192$, for
994 squares.

\vfill \eject
%\end
\hbox{}
\vs-.3 \nd
\epsfig .7\hsize; 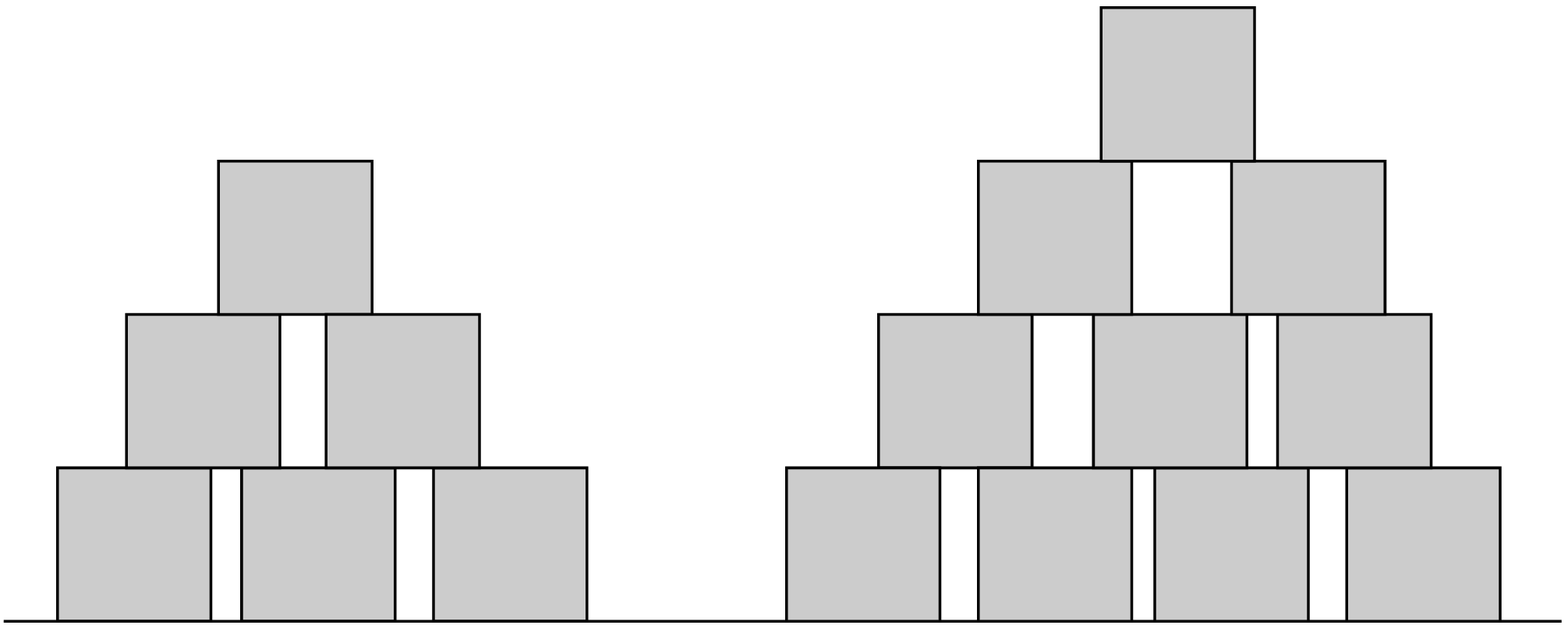;
\vs0
\centerline{Figure 3. An allowed configuration}

\vs0
\hbox{}
\vs.2 \nd
\epsfig .5\hsize; 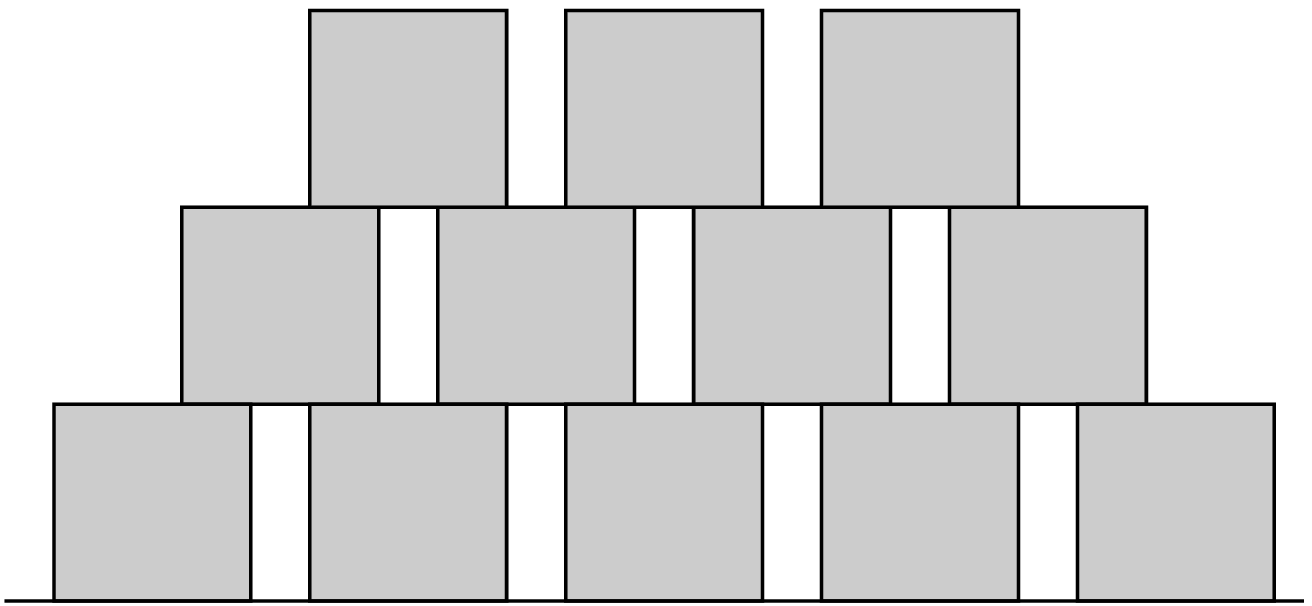;
\vs0
\centerline{Figure 4. A uniform configuration}
\vs0
\hbox{}
\vs0 \nd
\epsfig .85\hsize; 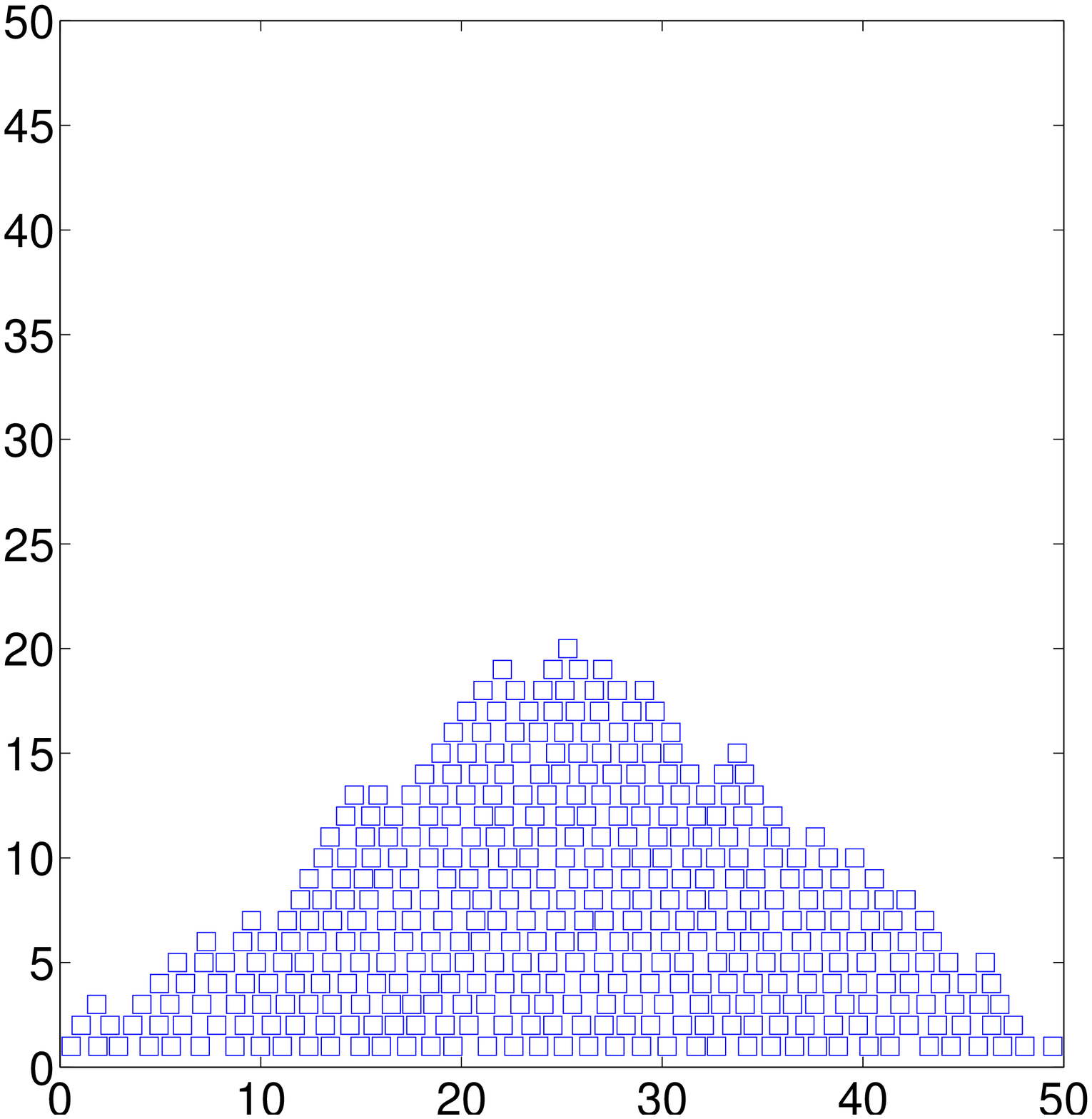; %600_8-config.eps;
\vs0
\centerline{Figure 5. 399 squares after $10^6$ moves.}

\vfill \eject
\hbox{}
\vs0 \nd
\epsfig 1\hsize; 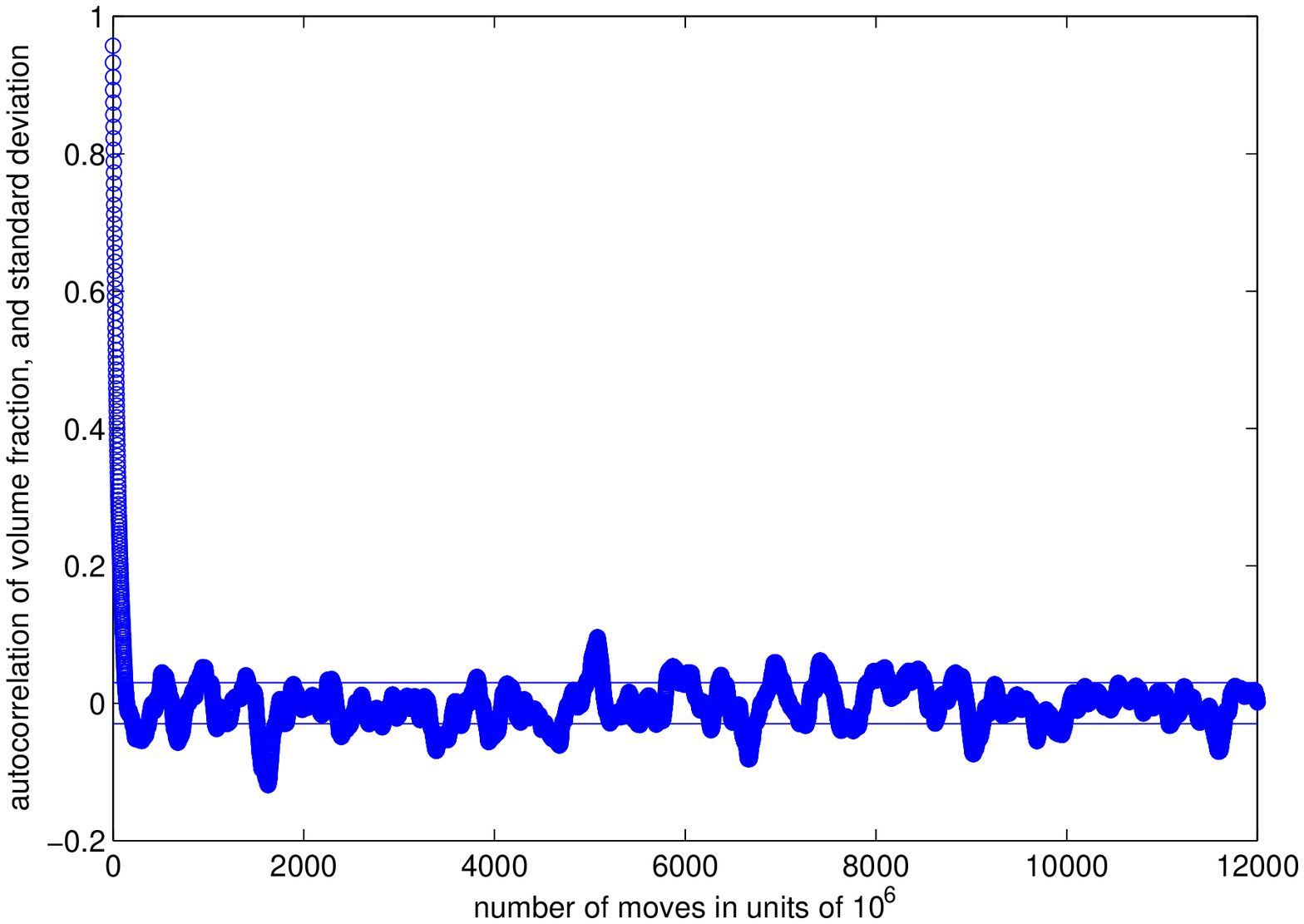;
\vs-.1 \nd
Figure 6a. Plot of biased autocorrelation (for 8000 particles) versus number of moves, with horizontal lines denoting one standard deviation away from zero.

\vs.2 \nd
\epsfig 1\hsize; 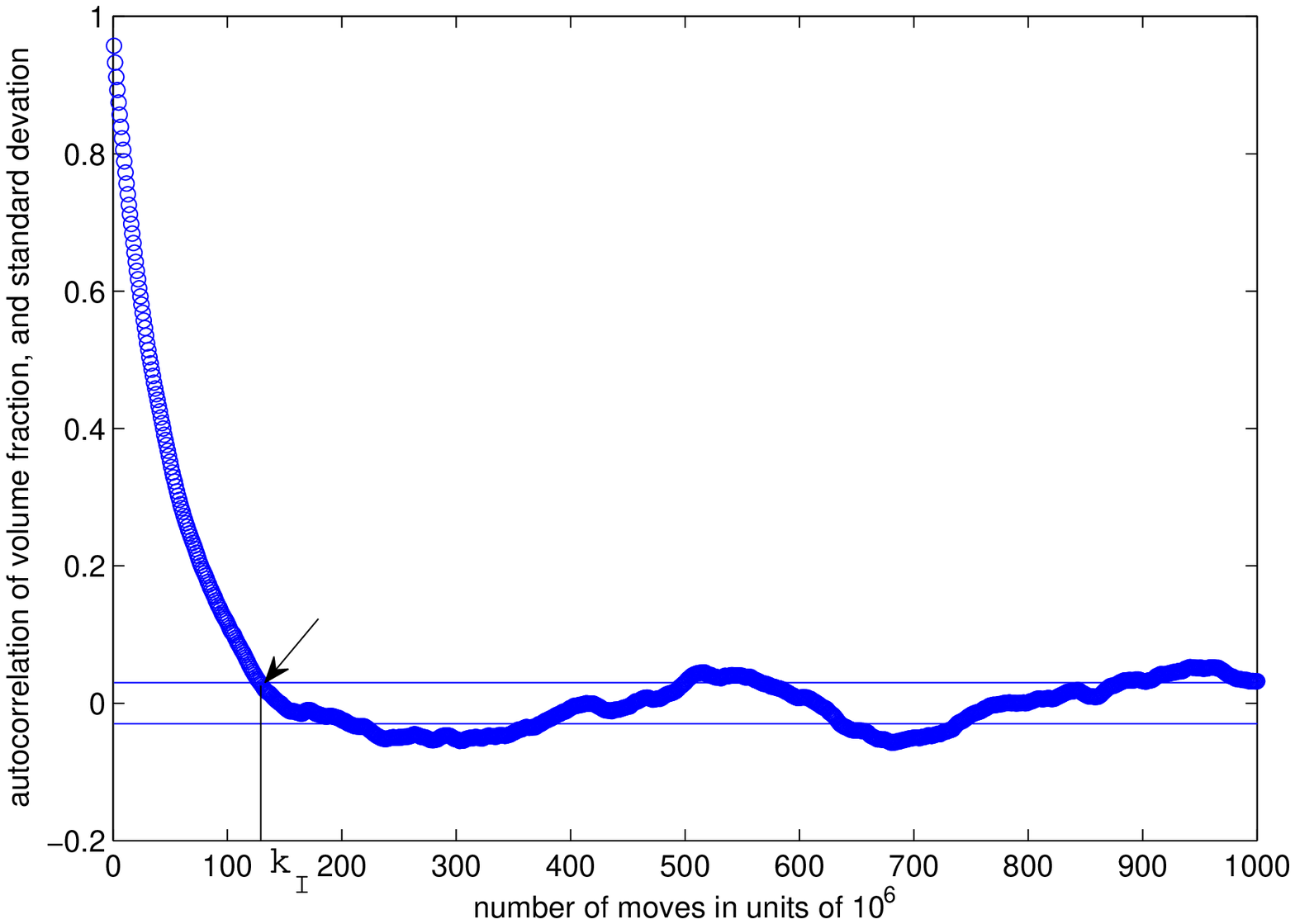;
\vs-.1 \nd
Figure 6b. Close-up of the above plot, with initialization time.

\vfill \eject
%\end
\hbox{}
\vs0 \nd
\epsfig 1.3\hsize; 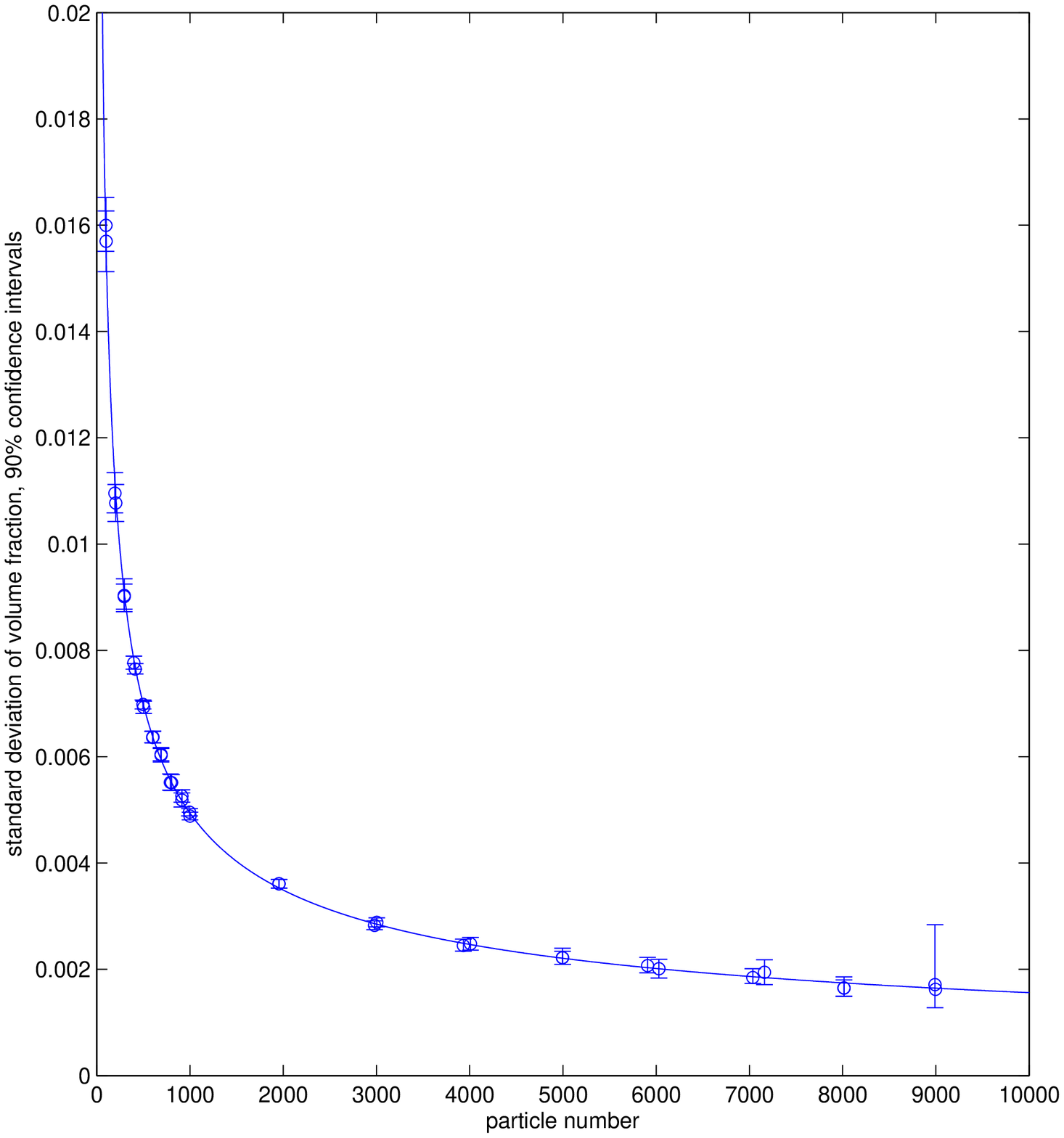;
\vs0 \nd
Figure 7. Plot of the standard deviation of the volume fraction versus
number of squares, using $f_2$ for confidence intervals.
\vfill \eject
%\end
\hbox{}
\vs0 \nd
\epsfig 1.3\hsize; 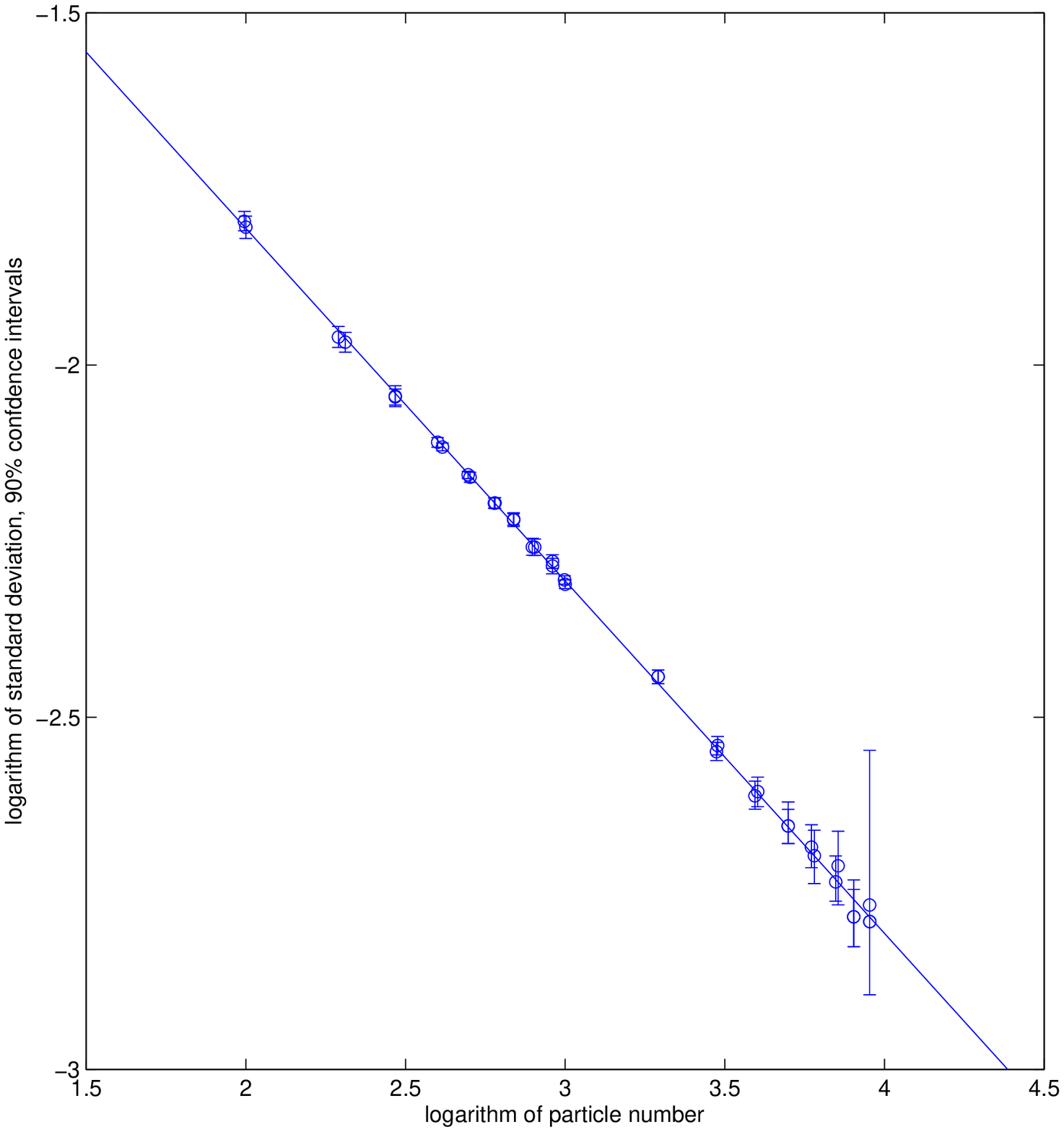;
\vs0 \nd
Figure 8. Plot of the standard deviation of the volume fraction versus
number of squares, using log scales and $f_2$ for confidence
intervals. The line is $y=-0.5004\,x-0.8052$.
\vfill \eject
%\end
\hbox{}
\vs0 \nd
\epsfig 1.3\hsize; 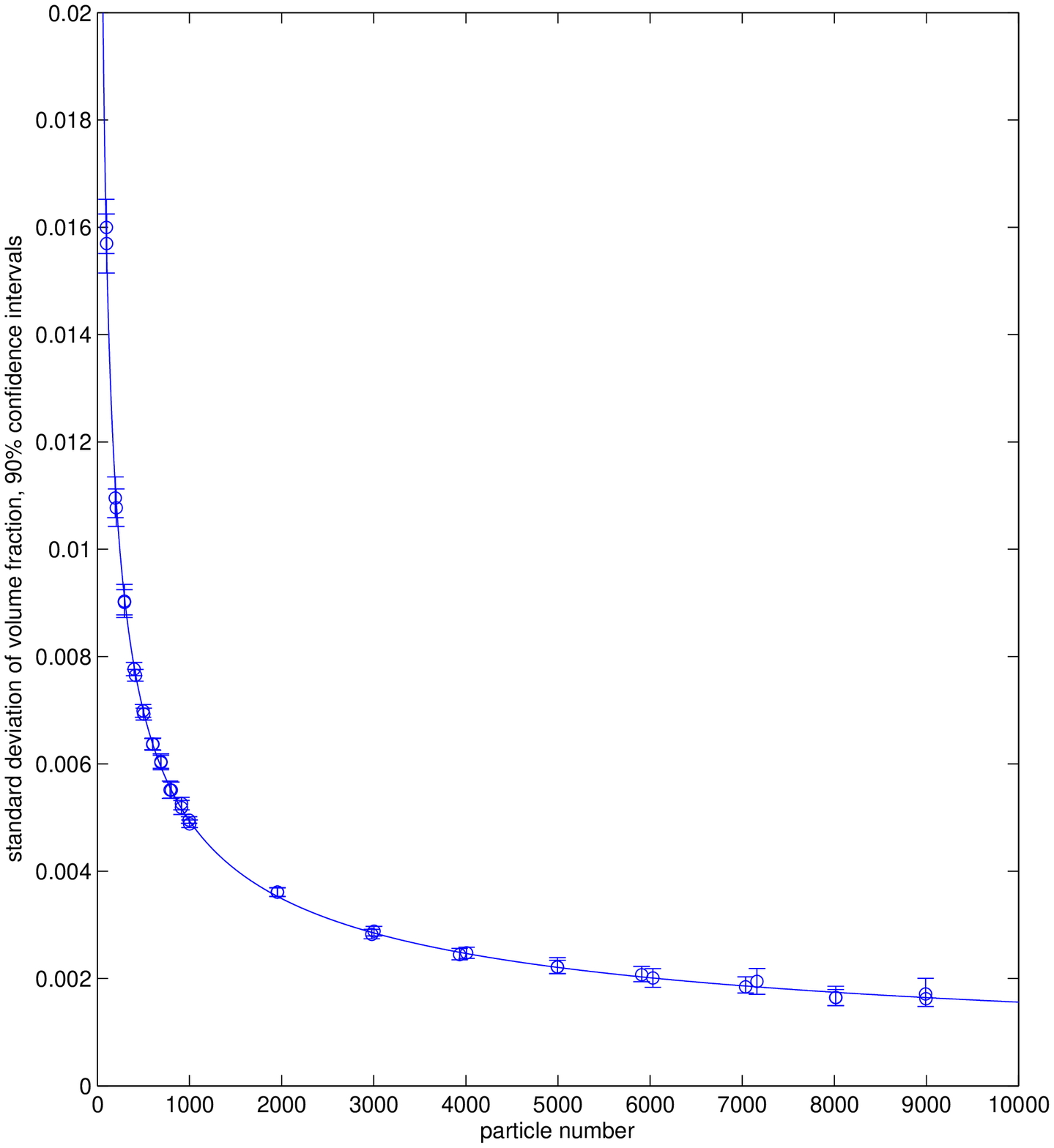;
\vs0 \nd
Figure 9. Plot of the standard deviation of the volume fraction versus
number of squares, using $f_1$ for confidence intervals.
\vfill \eject
%\end
\hbox{}
\vs0 \nd
\epsfig 1.3\hsize; 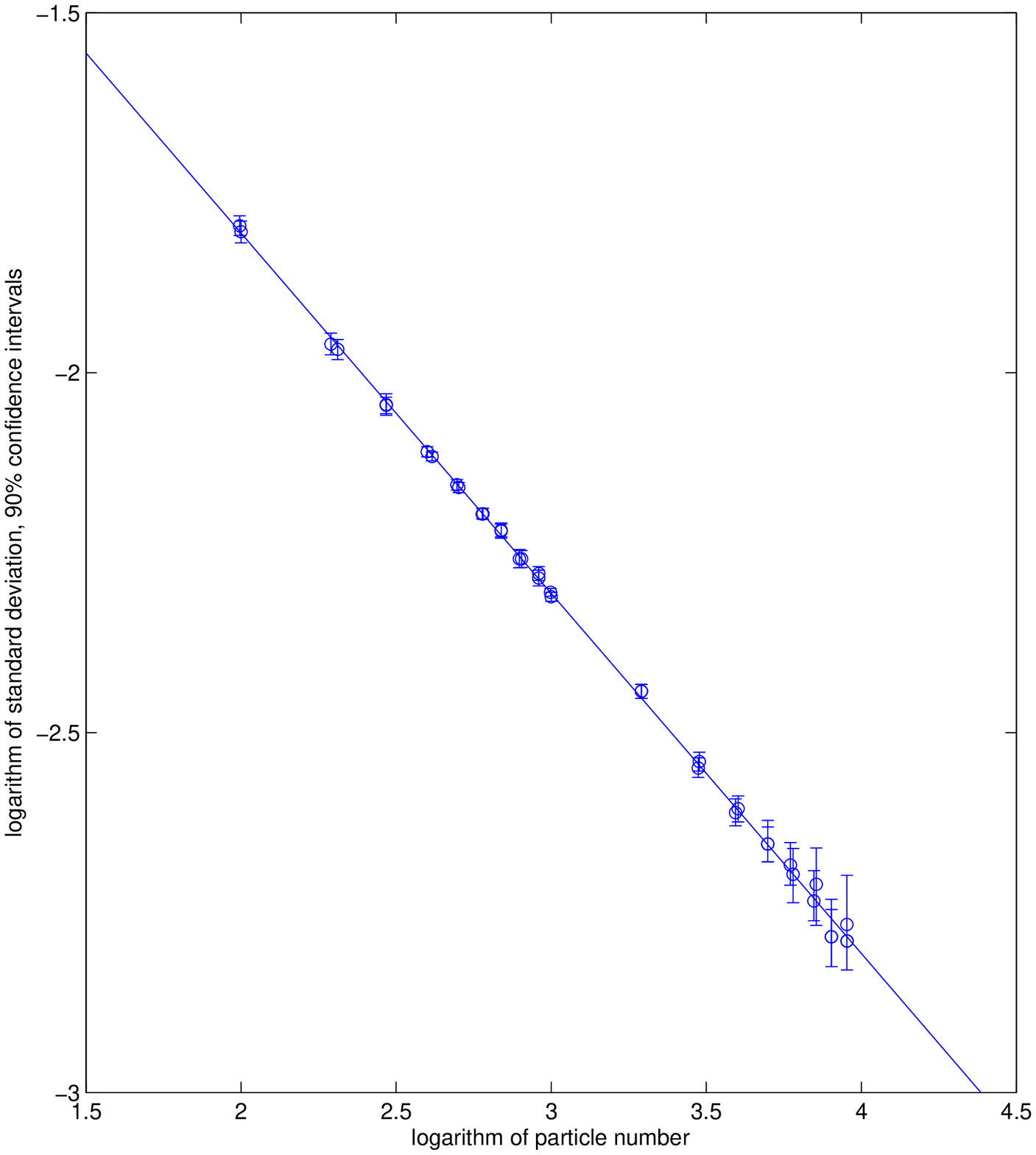;
\vs0 \nd
Figure 10. Plot of the standard deviation of the volume fraction versus
number of squares, using log scales and $f_1$ for confidence
intervals. The line is $y=-0.5003\,x-0.8055$.

\vfill \eject
%\end
\hbox{}
\vs0 \nd
\epsfig 1.3\hsize; 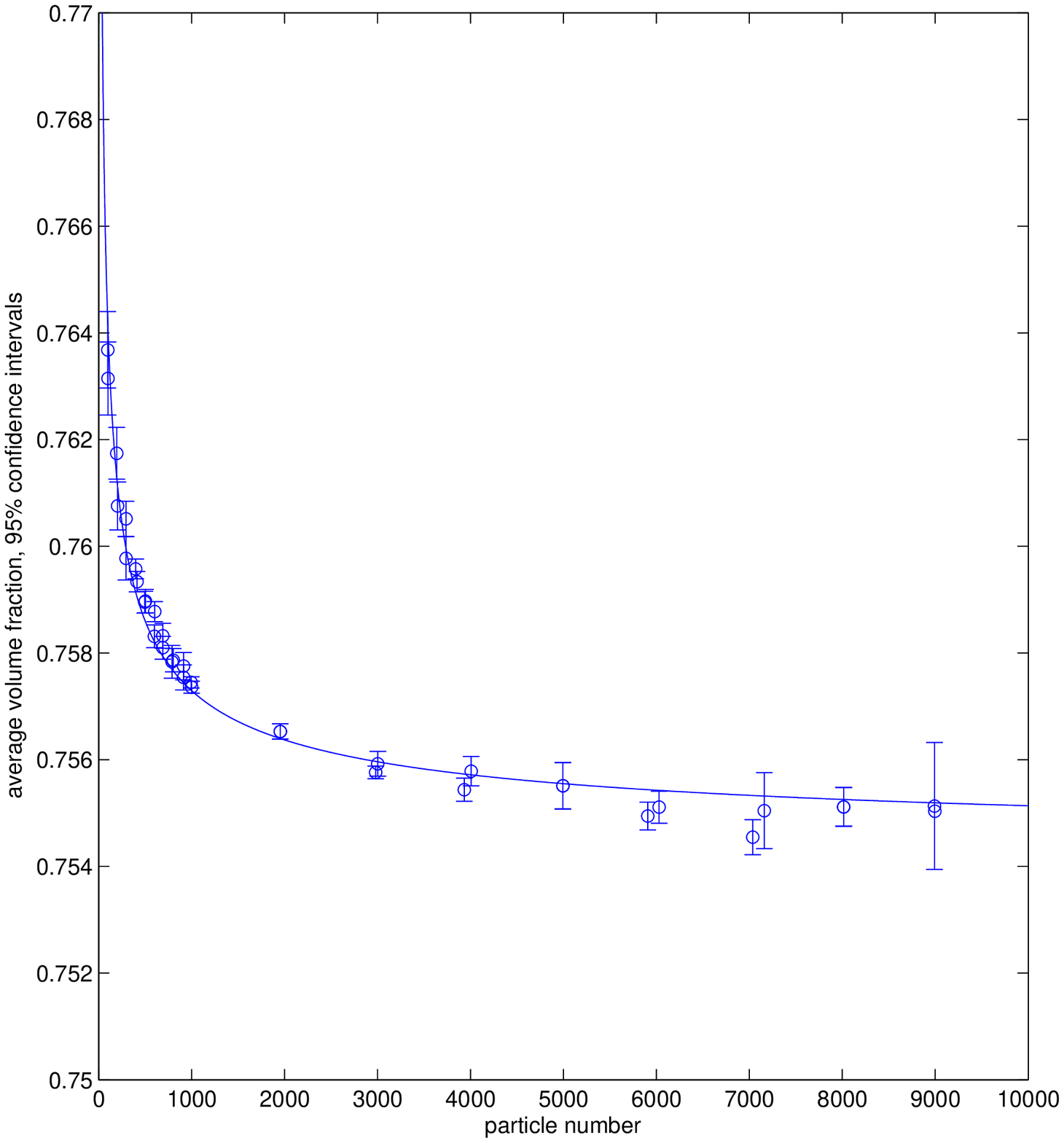;
\vs0 \nd
Figure 11. Plot of the mean of the volume fraction versus number of
squares, 
using $f_2$ for confidence intervals. The curve is $y=0.7541+0.0998\,x^{-1/2}$.
\vfill \eject
%\end
\hbox{}
\vs0 \nd
\epsfig 1.3\hsize; 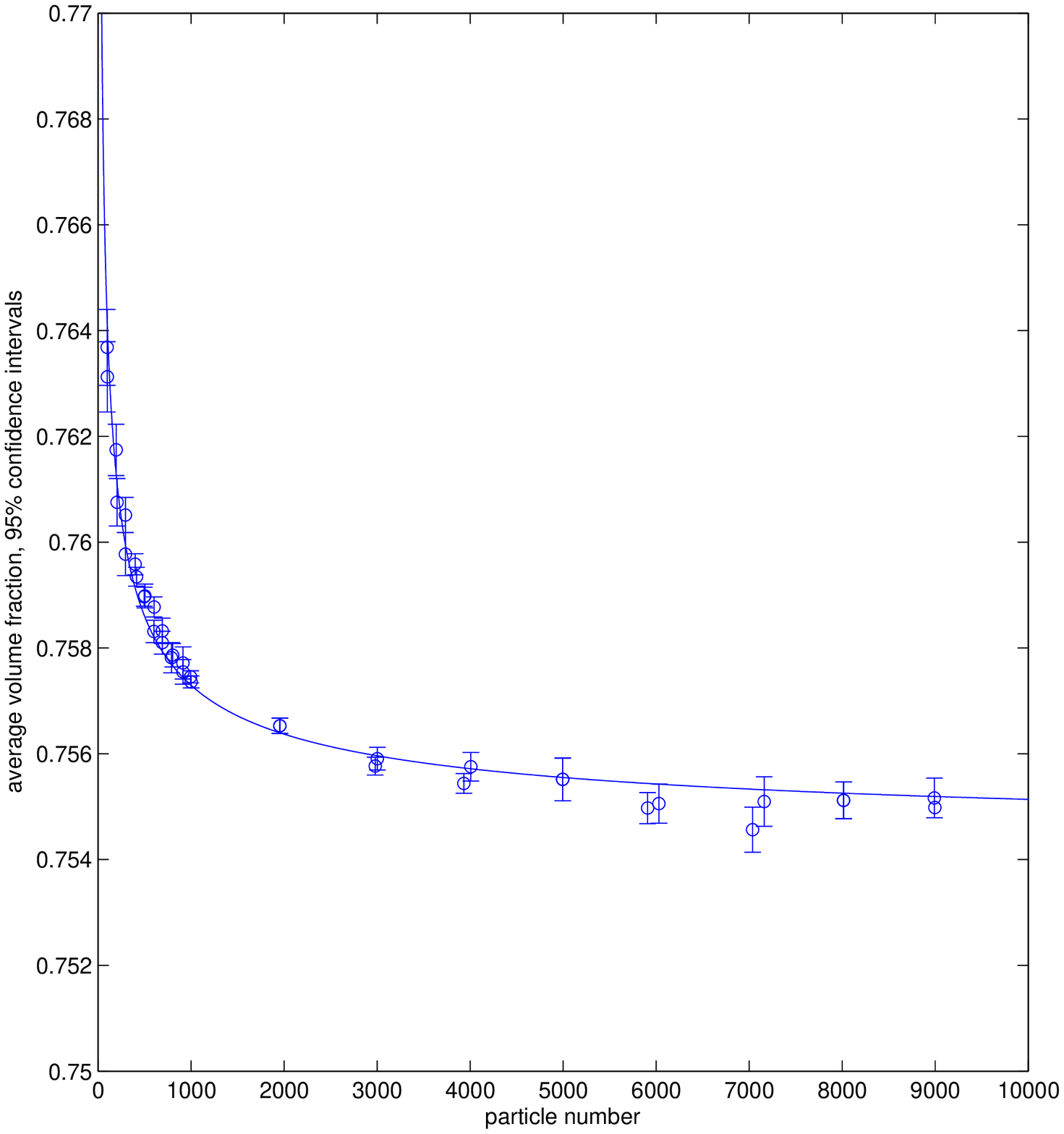;
\vs0 \nd
Figure 12. Plot of the mean of the volume fraction versus number of
squares, 
using $f_1$ for confidence intervals.  The curve is $y=0.7541+0.0998\,x^{-1/2}$.
\vfill \eject
%\end
\hbox{}
\vs0 \nd
\centerline{Table 1}
\vs.2
\centerline{Basics}
\vs.1
\centerline{(using unbiased autocorrelation $f_1$)}
\vs.4
{\ninepoint
\centerline{\vbox{
\halign{\hfil\indent#\hfil&\hfil\quad#\hfil&\hfil\quad#\hfil&\hfil\quad#\hfil&\hfil\quad#\hfil&\hfil\quad#\hfil\cr
number of &number of         &step size&$k_I$         &$k_M$         &run length\cr
squares   &moves in          &in units &(init. time)  &(mixing time) &in units of\cr
in packing&units of $10^{9}$ &of $10^6$&in units of   &in units of   &mixing time\cr
          &                  &moves    &step size     &step size     &using $f_1$\cr
          &                  &         &using $f_1$   &using $f_1$   &\cr
&&&&&\cr
99  &0.4&0.2&1  &1  &1998\cr
100 &0.4&0.2&1  &1  &1998\cr
195 &0.4&0.2&1  &1  &1998\cr
205 &0.4&0.2&1  &1  &1998\cr
294 &0.4&0.2&5  &5  &398 \cr
294 &0.4&0.2&4  &4  &498 \cr
399 &2  &0.2&10 &10 &998 \cr
413 &2  &0.2&9  &9  &1110\cr
497 &2  &0.2&17 &17 &587 \cr
504 &2  &0.2&17 &17 &587 \cr
600 &2  &0.2&12 &12 &832 \cr
603 &2  &0.2&23 &23 &433 \cr
690 &2  &0.2&14 &14 &713 \cr
690 &2  &0.2&20 &20 &498 \cr
790 &2  &0.2&43 &43 &231 \cr
803 &2  &0.2&14 &14 &713 \cr
913 &2  &0.2&34 &34 &293 \cr
913 &2  &0.2&75 &73 &135 \cr
996 &12 &1  &13 &13 &953 \cr
1001&12 &1  &16 &16 &755 \cr
1955&12 &1  &38 &38 &318 \cr
2980&12 &1  &44 &44 &277 \cr
3003&12 &1  &51 &53 &234 \cr
3933&12 &1  &64 &63 &193 \cr
4008&12 &1  &99 &75 &158 \cr
4995&12 &1  &193&174&68  \cr
5908&12 &1  &163&96 &125 \cr
6030&12 &1  &143&143&84  \cr
7037&12 &1  &223&261&46  \cr
7161&12 &1  &222&181&48  \cr
8015&12 &1  &132&120&100 \cr
8991&12 &1  &283&287&41  \cr
8995&12 &1  &632&631&18  \cr
}
}
}
}
\vfill \eject
\hbox{}
\vs0 \nd
\centerline{Table 2}
\vs.2
\centerline{Basics} 
\vs.1
\centerline{(using biased autocorrelation $f_2$)}
\vs.4
{\ninepoint
\centerline{\vbox{
\halign{\hfil\indent#\hfil&\hfil\quad#\hfil&\hfil\quad#\hfil&\hfil\quad#\hfil&\hfil\quad#\hfil&\hfil\quad#\hfil\cr
number of &number of         &step size&$k_I$         &$k_M$         &run length\cr
squares   &moves in          &in units &(init. time)  &(mixing time) &in units of\cr
in packing&units of $10^{9}$ &of $10^6$&in units of   &in units of   &mixing time\cr
          &                  &moves    &step size     &step size     &using $f_1$\cr
          &                  &         &using $f_2$   &using $f_2$   &\cr
&&&&&\cr
99  &0.4&0.2&1  &1  &1998\cr
100 &0.4&0.2&3  &2  &998 \cr
195 &0.4&0.2&1  &1  &1998\cr
205 &0.4&0.2&1  &1  &1998\cr
294 &0.4&0.2&5  &5  &398 \cr
294 &0.4&0.2&4  &4  &498 \cr
399 &2  &0.2&12 &12 &832 \cr
413 &2  &0.2&11 &11 &908 \cr
497 &2  &0.2&20 &20 &498 \cr
504 &2  &0.2&19 &19 &525 \cr
600 &2  &0.2&12 &12 &832 \cr
603 &2  &0.2&23 &23 &433 \cr
690 &2  &0.2&14 &15 &665 \cr
690 &2  &0.2&25 &20 &498 \cr
790 &2  &0.2&43 &44 &226 \cr
803 &2  &0.2&15 &14 &713 \cr
913 &2  &0.2&36 &38 &262 \cr
913 &2  &0.2&77 &75 &132 \cr
996 &12 &1  &18 &18 &687 \cr
1001&12 &1  &16 &16 &755 \cr
1955&12 &1  &38 &38 &318 \cr
2980&12 &1  &56 &56 &218 \cr
3003&12 &1  &51 &54 &230 \cr
3933&12 &1  &76 &78 &156 \cr
4008&12 &1  &101&88 &135 \cr
4995&12 &1  &191&177&67  \cr
5908&12 &1  &179&100&120 \cr
6030&12 &1  &139&151&80  \cr
7037&12 &1  &220&290&41  \cr
7161&12 &1  &199&186&47  \cr
8015&12 &1  &131&126&95  \cr
8991&12 &1  &282&355&33  \cr
8995&12 &1  &565&609&19  \cr
}
}
}
}
\vfill \eject

\hbox{}
\vs0 \nd
\centerline{Table 3}
\vs.2
\centerline{Volume Fraction}
\vs.4
{\ninepoint
\centerline{\vbox{
\halign{\hfil\indent#\hfil&\hfil\quad#\hfil&\hfil\quad#\hfil&\hfil\quad#\hfil\cr
number of &sample value &end points of     &end points of     \cr
squares   &             &$95\%$ confidence &$95\%$ confidence \cr
in packing&             &interval for      &interval for      \cr
          &             &true value,       &true value,       \cr
          &             &using $f_1$       &using $f_2$       \cr
&&&\cr
99  &0.7637&0.7637$\pm$0.0007&0.7637$\pm$0.0007\cr
100 &0.7631&0.7631$\pm$0.0007&0.7631$\pm$0.0007\cr
195 &0.7617&0.7617$\pm$0.0005&0.7617$\pm$0.0005\cr
205 &0.7608&0.7608$\pm$0.0005&0.7608$\pm$0.0005\cr
294 &0.7605&0.7605$\pm$0.0004&0.7605$\pm$0.0004\cr
294 &0.7598&0.7598$\pm$0.0004&0.7598$\pm$0.0004\cr
399 &0.7596&0.7596$\pm$0.0002&0.7596$\pm$0.0002\cr
413 &0.7593&0.7593$\pm$0.0002&0.7593$\pm$0.0002\cr
497 &0.7590&0.7590$\pm$0.0002&0.7590$\pm$0.0002\cr
504 &0.7590&0.7590$\pm$0.0003&0.7590$\pm$0.0002\cr
600 &0.7583&0.7583$\pm$0.0002&0.7583$\pm$0.0002\cr
603 &0.7588&0.7588$\pm$0.0002&0.7588$\pm$0.0002\cr
690 &0.7581&0.7581$\pm$0.0002&0.7581$\pm$0.0002\cr
690 &0.7583&0.7583$\pm$0.0003&0.7583$\pm$0.0003\cr
790 &0.7578&0.7578$\pm$0.0003&0.7578$\pm$0.0003\cr
803 &0.7579&0.7579$\pm$0.0002&0.7579$\pm$0.0002\cr
913 &0.7575&0.7575$\pm$0.0002&0.7575$\pm$0.0003\cr
913 &0.7578&0.7578$\pm$0.0003&0.7578$\pm$0.0003\cr
996 &0.7575&0.7575$\pm$0.0001&0.7575$\pm$0.0001\cr
1001&0.7574&0.7574$\pm$0.0001&0.7574$\pm$0.0001\cr
1955&0.7565&0.7565$\pm$0.0002&0.7565$\pm$0.0002\cr
2980&0.7558&0.7558$\pm$0.0002&0.7558$\pm$0.0001\cr
3003&0.7559&0.7559$\pm$0.0002&0.7559$\pm$0.0003\cr
3933&0.7554&0.7554$\pm$0.0002&0.7554$\pm$0.0002\cr
4008&0.7558&0.7558$\pm$0.0003&0.7558$\pm$0.0003\cr
4995&0.7555&0.7555$\pm$0.0004&0.7555$\pm$0.0005\cr
5908&0.7549&0.7549$\pm$0.0003&0.7549$\pm$0.0003\cr
6030&0.7551&0.7551$\pm$0.0004&0.7551$\pm$0.0003\cr
7037&0.7545&0.7545$\pm$0.0005&0.7545$\pm$0.0004\cr
7161&0.7550&0.7550$\pm$0.0005&0.7550$\pm$0.0007\cr
8015&0.7551&0.7551$\pm$0.0004&0.7551$\pm$0.0004\cr
8991&0.7551&0.7551$\pm$0.0004&0.7551$\pm$0.0012\cr
8995&0.7550&none             &none\cr
}
}
}
}
\vfill \eject

\hbox{}
\vs0 \nd
\centerline{Table 4}
\vs.2
\centerline{Standard deviation of volume fraction}
\vs.4
{\ninepoint
\centerline{\vbox{
\halign{\hfil\indent#\hfil&\hfil\quad#\hfil&\hfil\quad#\hfil&\hfil\quad#\hfil\cr
number of &sample value &end points of     &end points of     \cr
squares   &             &$90\%$ confidence &$90\%$ confidence \cr
in packing&             &interval for      &interval for      \cr
          &             &true value,       &true value,       \cr
          &             &using $f_1$       &using $f_2$       \cr
&&&\cr
99  &0.0160&0.0160$\pm$0.0005&0.0160$\pm$0.0005\cr
100 &0.0157&0.0157$\pm$0.0006&0.0157$\pm$0.0006\cr
195 &0.0110&0.0110$\pm$0.0004&0.0110$\pm$0.0004\cr
205 &0.0108&0.0108$\pm$0.0004&0.0108$\pm$0.0004\cr
294 &0.0090&0.0090$\pm$0.0003&0.0090$\pm$0.0003\cr
294 &0.0090&0.0090$\pm$0.0003&0.0090$\pm$0.0003\cr
399 &0.0078&0.0078$\pm$0.0001&0.0078$\pm$0.0002\cr
413 &0.0076&0.0077$\pm$0.0001&0.0077$\pm$0.0001\cr
497 &0.0070&0.0070$\pm$0.0001&0.0070$\pm$0.0001\cr
504 &0.0069&0.0069$\pm$0.0001&0.0069$\pm$0.0001\cr
600 &0.0064&0.0064$\pm$0.0001&0.0064$\pm$0.0001\cr
603 &0.0064&0.0064$\pm$0.0001&0.0064$\pm$0.0001\cr
690 &0.0060&0.0061$\pm$0.0002&0.0060$\pm$0.0001\cr
690 &0.0060&0.0060$\pm$0.0002&0.0060$\pm$0.0001\cr
790 &0.0055&0.0055$\pm$0.0002&0.0055$\pm$0.0002\cr
803 &0.0055&0.0055$\pm$0.0002&0.0055$\pm$0.0002\cr
913 &0.0052&0.0052$\pm$0.0002&0.0052$\pm$0.0002\cr
913 &0.0053&0.0053$\pm$0.0001&0.0053$\pm$0.0001\cr
996 &0.0050&0.0050$\pm$0.0001&0.0050$\pm$0.0001\cr
1001&0.0049&0.0049$\pm$0.0001&0.0049$\pm$0.0001\cr
1955&0.0036&0.0036$\pm$0.0001&0.0036$\pm$0.0001\cr
2980&0.0028&0.0028$\pm$0.0001&0.0028$\pm$0.0001\cr
3003&0.0029&0.0029$\pm$0.0001&0.0029$\pm$0.0001\cr
3933&0.0024&0.0025$\pm$0.0001&0.0025$\pm$0.0001\cr
4008&0.0025&0.0025$\pm$0.0001&0.0025$\pm$0.0002\cr
4995&0.0022&0.0022$\pm$0.0002&0.0022$\pm$0.0002\cr
5908&0.0021&0.0021$\pm$0.0002&0.0021$\pm$0.0002\cr
6030&0.0020&0.0020$\pm$0.0002&0.0020$\pm$0.0002\cr
7037&0.0018&0.0019$\pm$0.0002&0.0019$\pm$0.0002\cr
7161&0.0019&0.0020$\pm$0.0003&0.0021$\pm$0.0004\cr
8015&0.0016&0.0017$\pm$0.0002&0.0017$\pm$0.0002\cr
8991&0.0017&0.0017$\pm$0.0003&0.0021$\pm$0.0008\cr
8995&0.0016&none             &none             \cr
}
}
}
}
\vfill \eject

\hbox{}
\vs0 \nd
\centerline{Table 5}
\vs.2
\centerline{Fraction of times the given batch size gives acceptable confidence
interval for}
\centerline{given segment of total data of long runs, using
unbiased autocorrelation $f_1$}

\vs.5
{\ninepoint
\centerline{\vbox{\offinterlineskip
\halign{\strut
\indent#\hfil
&\hfil\quad#\hfil
&\hfil\quad#\enspace\hfil
&\vrule#
&\hfil\enspace#\hfil
&\hfil\quad#\hfil
&\hfil\quad#\hfil
&\hfil\quad#\hfil
&\hfil\quad#\hfil
&\hfil\quad#\hfil\cr
&&&\omit&20-100&100-200&200-300&300-400&400-500&500-600\cr
&&&\omit&mixing&mixing&mixing&mixing&mixing&mixing\cr
&&&\omit&times of&times of&times of&times of&times of&times of\cr
&&&\omit&total data&total data&total data&total data&total data&total data\cr
&&&\omit&\multispan6\hrulefill\cr
\noalign{\vskip-3pt}             % adjust if lines do not meet correctly
&&&&&&&&&\cr
&&1-5&&0.0849&0.0867&0.1119&0.1191&0.1938&0.2082\cr
number of&&6-10&&0.9410&0.9394&0.9830&1.0000&1.0000&1.0000\cr
mixing&&11-15&&0.9648&0.9231& 0.9656&1.0000&1.0000&1.0000\cr
times&&16-20&&0.9524&0.9095&0.9777&0.9879&1.0000&1.0000\cr
per batch&&21-31&&0.9650&0.9177&0.9673&1.0000&1.0000&1.0000\cr
&&31-40&&0.9712&0.9042&0.9643&1.0000&1.0000&0.9957\cr
&&41-51&&0.9375&0.8869&0.9402&0.9511&0.9783&0.9402\cr
}
}
}
}

\vs1 \nd
\centerline{Table 6}
\vs.2
\centerline{Fraction of times the given batch size gives acceptable confidence
interval for}
\centerline{given segment of total data of long runs, using biased
autocorrelation $f_2$}
\vs.5
{\ninepoint
\centerline{\vbox{\offinterlineskip
\halign{\strut
\indent#\hfil
&\hfil\quad#\hfil
&\hfil\quad#\enspace\hfil
&\vrule#
&\hfil\enspace#\hfil
&\hfil\quad#\hfil
&\hfil\quad#\hfil
&\hfil\quad#\hfil
&\hfil\quad#\hfil
&\hfil\quad#\hfil\cr
&&&\omit&20-100&100-200&200-300&300-400&400-500&500-600\cr
&&&\omit&mixing&mixing&mixing&mixing&mixing&mixing\cr
&&&\omit&times of&times of&times of&times of&times of&times of\cr
&&&\omit&total data&total data&total data&total data&total data&total data\cr
&&&\omit&\multispan6\hrulefill\cr
\noalign{\vskip-3pt}             % adjust if lines do not meet correctly
&&&&&&&&&\cr
&&1-5&&0.0833&0.0924&0.1182&0.1259&0.2006&0.2326\cr
number of&&6-10&&0.9245&0.9784&0.9805&0.9552&0.9762&1.0000\cr
mixing&&11-15&&0.9107&0.9451&0.9607&0.9524&0.9707&1.0000\cr
times&&16-20&&0.9728&0.9212&0.9745&0.9493&0.9655&1.0000\cr
per batch&&21-31&&0.9486&0.9059&0.9592&0.9547&0.9744&1.0000\cr
&&31-40&&0.9780&0.9190&0.9592& 0.9704&0.9852&1.0000\cr
&&41-51&&0.9000&0.8639&0.9441&0.9565&0.9814&1.0000\cr
}
}
}
}
\vfill \eject

\hbox{} \nd
{\bf Acknowledgements}. We gratefully acknowledge useful discussions
with
P. Diaconis, W.D. McCormick, M. Schr\"oter and H.L. Swinney.
\vs.2
\centerline{\bf References}
\vs.2
\item{[AH]}
B.J. Alder and W.G. Hoover, Numerical Statistical Mechanics, in Physics
of Simple Liquids, edited by H.N.V. Temperley, J.S. Rowlinson and
G.S. Rushbrooke, (John Wiley, New York, 1968) 79-113.
\item{[DL]} P. Diaconis and E. Lehmann, Comment,
J. Amer. Stat. Assoc. 103 (2008) 16-19.
\item{[dG]} P.G. de Gennes, Granular matter: a tentative view. Rev. Mod.
  Phys. {71} (1999) S374--S382.
\item{[EO]} S.F. Edwards and
  R.B.S. Oakeshott, Theory of powders, Physica A 157 (1989) 1080-1090.
\item{[Gey]} C.J. Geyer, Practical Markov chain Monte Carlo,
Stat. Sci. 7 (1992) 473-483.
\item{[Gel]} A. Gelman et al, 
Stat. Sci. 7 (1992) 457-511.
\item{[H]} W.G. Hoover, 
Bounds on the configurational integral for hard parallel squares and cubes,
J. Chem. Phys. 43 (1965) 371-374.
\item{[JS]} M. Jerkins, M. Schr\"oter, H.L. Swinney, T.J. Senden,
M. Saadatfar and T. Aste, 
Onset of mechanical stability in random packings of frictional 
particles,
Phys. Rev. Lett. 101 (2008) 018301.
\item{[KV]} C. Kipnis and S.R.S. Varadhan, 
Central limit theorem for additive functionals of reversible Markov
processes and applications to simple exclusions, Commun. math. phys. 104
(1986) 1-19.
\item{[MP]} R. Monasson and O. Pouliquen, 
Entropy of particle packings: an illustration on a toy model,
{Physica A} 236 (1997) 395-410.
\item{[NB]} 
M.E.J. Newman and G.T. Barkema, Monte Carlo methods in statistical
physics, (Oxford University Press, 1999).
\item{[P]} M.B. Priestley, Spectral analysis and time series, vol.1,
(Academic Press, New York, 1981).
\item{[PC]} M. Pica Ciamarra, A. Coniglio,
Random very loose packs, cond-mat/0805.0220
\item{[R]} C. Radin, Random close packing of granular matter, 
J. Stat. Phys. 131 (2008) 567-573.
\item{[S]} G.D. Scott,
  Packing of spheres, Nature (London) 188 (1960) 908-909.  
\item{[SK]} G.D. Scott and D.M. Kilgour, The
  density of random close packing of spheres,
  Brit. J. Appl. Phys. (J. Phys. D) 2 (1969) 863-866.  
\item{[SN]} M. Schr\"oter, S. N\"agle, C. Radin and H.L. Swinney,
  Phase transition in a static granular system, Europhys. Lett. 78
  (2007) 44004.  
\item{[TT]} S. Torquato, T.M. Truskett and
  P.G. Debenedetti, Is random close packing of spheres well defined?,
  Phys. Rev. Lett. 84 (2000) 2064-2067.  

\end